\documentclass[a4paper,11pt]{article}
\pdfoutput=1
\usepackage{jheppub}

\usepackage{graphicx}
\usepackage[figuresright]{rotating}
\usepackage{bm,amsmath,amssymb}
\usepackage[mathscr]{eucal}
\usepackage{multirow}
\usepackage{xcolor}
\usepackage{mathrsfs}
\usepackage{mathtools}
\usepackage{slashed}
\usepackage{graphics}
\usepackage{graphicx}
\usepackage{subfigure}
\usepackage{dsfont}
\usepackage{longtable}
\usepackage{bbm} 
\usepackage{float}
\usepackage{tcolorbox}
\usepackage{lipsum}
\usepackage{cancel}
\usepackage{enumerate}
\usepackage{makecell}

\definecolor{lcolor}{rgb}{0.,0.0,0.}
\definecolor{citcolor}{rgb}{0,0.,0.5}

\newcommand{\rmL}{\textrm{L}}
\newcommand{\rmT}{\textrm{T}}
\newcommand{\Hcal}{\mathcal{H}}

\newcommand{\Scal}{\mathcal{S}}

\newcommand{\vect}[1]{\boldsymbol{#1}_{\perp}}

\newcommand{\kgt}{\boldsymbol{k_{g\perp}}}

\newcommand{\Pt}{\vect{P}}

\newcommand{\Pthat}{\boldsymbol{\hat P}_{\perp}}

\newcommand{\at}{\vect{a}}
\newcommand{\bt}{\vect{b}}

\newcommand{\Kt}{\vect{K}}

\newcommand{\ruupt}{\boldsymbol{r}_{uu'}}
\newcommand{\ktone}{\boldsymbol{k_{1\perp}}}
\newcommand{\kttwo}{\boldsymbol{k_{2\perp}}}
\newcommand{\ktoneh}{\boldsymbol{k_{h1\perp}}}
\newcommand{\kttwoh}{\boldsymbol{k_{h2\perp}}}

\newcommand{\xt}{\vect{x}}
\newcommand{\yt}{\vect{y}}

\newcommand{\ut}{\vect{u}}

\newcommand{\rt}{\vect{r}}

\newcommand{\Lt}{\vect{L}}

\newcommand{\rxyt}{\boldsymbol{r}_{xy}}
\newcommand{\rbbpt}{\boldsymbol{r}_{bb'}}

\newcommand{\rxxtp}{\boldsymbol{r}_{xx'}}
\newcommand{\ryytp}{\boldsymbol{r}_{yy'}}

\newcommand{\der}{\mathrm{d}}

\newcommand{\Tr}{\mathrm{Tr}}

\begin{document}

\title{Dihadron correlations in small-$x$ DIS at NLO: transverse momentum dependent fragmentation}
\author[a]{Paul Caucal,}
\emailAdd{caucal@subatech.in2p3.fr}
\author[b]{Farid Salazar }
\emailAdd{faridsal@uw.edu}
 \affiliation[a]{SUBATECH UMR 6457 (IMT Atlantique, Université de Nantes, IN2P3/CNRS), 4 rue Alfred Kastler, 44307 Nantes, France}
\affiliation[b]{Institute for Nuclear Theory, University of Washington, Seattle WA 98195-1550, USA}

\abstract{
We compute inclusive dihadron cross-section in Deep Inelastic Scattering at next-to-leading order (NLO) and small $x$ in the Color Glass Condensate. We focus on the kinematic limit where the hadrons are produced at forward rapidities (in the direction of the virtual photon) and back-to-back in the transverse plane. Our calculation demonstrates that the coefficient of the Sudakov double logarithm for this process is $-\frac{\alpha_s}{2\pi}\left[C_F+\frac{N_c}{2}\right]$ instead of $-\frac{\alpha_s N_c}{4\pi}$ when back-to-back jets are measured in the final state. To preserve the universality of the Sudakov soft factor associated with the Weizs\"{a}cker-Williams transverse momentum dependent (TMD) gluon distribution, we promote the collinear fragmentation functions into TMD fragmentation functions. We then perform the resummation of the Sudakov logarithms through Collins-Soper-Sterman evolution of the TMD fragmentation functions and the Weizs\"{a}cker-Williams TMD gluon distribution. Finally, analytic expressions are obtained for the NLO coefficient functions in the $\overline{\rm MS}$-scheme. These results pave the way towards numerically calculating dihadron correlations at small $x$ at the future Electron-Ion Collider with full NLO accuracy.
}

\maketitle

In the high energy limit of Deep Inelastic Scattering (DIS) of electrons off nuclei, the increase in the number of gluons in the hadron wavefunction is expected to be tamed by non-linear gluon recombination effects \cite{Gribov:1984tu,Mueller:1985wy}, a phenomenon known as gluon saturation \cite{McLerran:1993ka,McLerran:1993ni,Kovchegov:1996ty}. In this regime, hadronic matter reaches a universal state characterized by a high gluon occupancy. The Color Glass Condensate (CGC) is an effective theory to systematically investigate this regime of matter \cite{Iancu:2003xm,Gelis:2010nm,Kovchegov:2012mbw}. Among many experimental signatures (see e.g.\,~\cite{Morreale:2021pnn} for a review), such a universal gluon-saturated state is expected to leave its imprint through the suppression of two-particle azimuthal correlations at forward rapidities \cite{Kharzeev:2004bw,Marquet:2007vb,Dominguez:2010xd,Dominguez:2011wm} which has been observed in deuteron-gold \cite{PHENIX:2011puq} and proton-gold collisions at RHIC~\cite{STAR:2021fgw}. A similar suppression is expected in the production of dihadrons in nuclear DIS at the Electron-Ion Collider (EIC) at small-$x$~\cite{Zheng:2014vka}.

In the saturation/CGC formalism, the transverse momentum imbalance of the two-particle production in the forward direction is expected to be on the order of the nuclear-dependent saturation scale $Q_s$, resulting in a broadening of the azimuthal two-particle correlations around $\phi \sim \pi$. Furthermore, the small-$x$ evolution of the saturated gluon distribution suppresses the azimuthal correlations \cite{Kharzeev:2003wz}. On the other hand, soft gluon radiations at large angles play a predominant role in competing with the effect due to gluon saturation, insofar as they also wash away azimuthal correlations imprinted by gluon saturation. From a perturbative QCD perspective, the leading $\mathcal{O}(\alpha_s)$ corrections to inclusive dihadron production in DIS are enhanced by large double logarithms in the ratio of the hadron relative transverse momentum $\hat P_\perp$ over the dihadron transverse momentum imbalance $\hat K_\perp$. For $\alpha_s\ln^2(\hat P_\perp/\hat K_\perp)\sim 1$, an all-order resummation within a Sudakov soft factor is necessary to stabilize the perturbative expansion~\cite{Sudakov:1954sw,Dokshitzer:1978hw,Parisi:1979se,Collins:1984kg}.

Previous studies~\cite{Stasto:2018rci,Benic:2022ixp} of dihadron or gamma-hadron correlations at small $x$ employed LO results in the Color Glass Condensate (CGC) effective field theory, supplemented with a Sudakov soft factor borrowed from calculations usually performed in the context of transverse momentum dependent (TMD) collinear factorization~\cite{Collins:2011zzd}. While this has been motivated by NLO calculations at small-$x$ for color neutral or jet final states (Higgs or $\gamma$-jet production in $pA$ collisions, dijet production in DIS)~\cite{Mueller:2012uf,Mueller:2013wwa,Xiao:2017yya}, there is no, to the best of our knowledge, first principle calculation to justify the universality of the Sudakov double logarithms between the Bjorken and the Regge limit when hadrons instead of jets are measured in the final state.

In this paper, we thus consider for the first time the production of a pair of back-to-back (in the transverse plane) hadrons in DIS at small-$x$ at NLO in the CGC. At LO in the CGC and at the partonic level, this process factorizes in terms of the Weizs\"{a}cker-Williams (WW) gluon TMD distribution~\cite{Dominguez:2010xd,Dominguez:2011wm}, which represents, in light-cone gauge, the number density of gluons in the target. At NLO, it has recently been shown that this factorization still holds for jet-defined final states \cite{Caucal:2023fsf} with an explicit calculation of the Sudakov factor for the WW gluon TMD at single logarithmic accuracy (extending the results obtained in \cite{Mueller:2013wwa} for the Sudakov double logarithm of this process at small $x$). It is natural to ask whether this factorization remains for inclusive back-to-back dihadron production in DIS at NLO and to compute the associated leading Sudakov logarithms. The interest in looking at hadronic final states in DIS is twofold: (i) from a theoretical perspective, it enables one to clarify the connection between TMD factorization in the small-$x$ regime and TMD factorization in the collinear regime (ii) from a phenomenological point of view, hadron measurements at low $p_t$ (where saturation effects are enhanced) at the future EIC should be more accessible than dijet measurements \cite{Zheng:2014vka}. 

We therefore explicitly compute the NLO corrections in the back-to-back limit for this process and isolate the leading Sudakov double and single logarithms. As expected from the insight of transverse momentum-dependent factorization theorems in other DIS processes like SIDIS~\cite{Ji:2004wu,Ji:2004xq}, the Sudakov double logarithm we obtain in the case of inclusive dihadron production is different from the Sudakov double logarithm for inclusive dijet production. The difference is due to the fragmentation contribution of the outgoing quark and antiquark pair. Inspired by the Collins-Soper-Sterman (CSS) formalism \cite{Collins:1981uk,Collins:1981uw,Collins:1984kg,Collins:2011zzd}, we propose a factorized expression in terms of the WW gluon TMD and two quark/antiquark TMD fragmentation functions (instead of standard ``integrated" fragmentation functions) to consistently resum the large Sudakov logarithms to all orders. While this is a well-known procedure in the context of $k_\perp$-dependent collinear factorization (see e.g.~\cite{Boussarie:2023izj,Metz:2016swz} for reviews), it is the first time such a framework is shown to be valid at small $x$ to one loop order. As in the dijet case, the small-$x$ evolution of the WW gluon TMD is dictated by B-JIMWLK equation \cite{Balitsky:1995ub,JalilianMarian:1996xn,JalilianMarian:1997dw,Kovner:2000pt,Iancu:2000hn,Iancu:2001ad,Ferreiro:2001qy} amended with a kinematic constraint~\cite{Andersson:1995ju,Kwiecinski:1996td,Kwiecinski:1997ee,Salam:1998tj,Ciafaloni:1998iv,Ciafaloni:1999yw,Ciafaloni:2003rd,SabioVera:2005tiv,Motyka:2009gi,Beuf:2014uia,Iancu:2015vea,Ducloue:2019ezk}, which for the process at hand allows for the separation between small-$x$ gluons contributing to B-JIMWLK, and soft gluons contributing to the Sudakov~\cite{Taels:2022tza,Caucal:2022ulg}.

Finally, we compute in the $\overline{\rm MS}$ scheme the NLO coefficient functions, defined as the pure $\mathcal{O}(\alpha_s)$ corrections in the back-to-back limit (leading power in $K_\perp/P_\perp$) which are free of any large small-$x$ or Sudakov logarithms. As expected, these NLO coefficient functions also differ from the dijet case since coefficient functions are process-dependent. Combining these results, our factorized expression is built from three essential ingredients: (i) the CSS-evolved and small-$x$ evolved WW TMD, (ii) the Dokshitzer-Gribov-Lipatov-Altarelli-Parisi (DGLAP)~\cite{Gribov:1972ri,Dokshitzer:1977sg,Altarelli:1977zs} and CSS evolved TMD fragmentation functions of $q\bar q$ pair into hadrons, and (iii) the NLO coefficient function. The ingredient (ii) is the main difference between a dihadron and a dijet measurement. Our factorized expression should considerably simplify future phenomenological studies of back-to-back dihadron correlations at small $x$ at NLO at the EIC.

Our paper is structured as follows. In the first section, we briefly review the leading order result for inclusive dihadron cross-section in the back-to-back limit and introduce our notations for the kinematic variables. Section \ref{sec:coll-fact-FF} is dedicated to the proof of the factorization of final state collinear divergences in terms of fragmentation functions evolving via the DGLAP equation when the dihadron pair is produced with general small-$x$ kinematics (not necessarily back-to-back). The details of this rather standard calculation can be found in appendix~\ref{app:DGLAP-frag}. In section \ref{sec:heuristic-discussion}, we isolate the leading Sudakov logarithm emerging from the impact factor when the back-to-back limit is considered, and we discuss the difference between the dihadron and the dijet case and highlight their physical origin. In section~\ref{eq:NLO-if-dihadron}, we go beyond double logarithmic accuracy and explicitly compute the single Sudakov logarithm and finite pieces in the NLO impact factor. Section~\ref{sec:resum} deals with the resummation of the Sudakov double and single logarithms within a fully factorized expression involving the WW gluon TMD and quark/antiquark TMD fragmentation functions. Our final result for the inclusive dihadron production in DIS at small-$x$ at NLO in the back-to-back limit is presented in this section. We conclude in section~\ref{sec:outlook} with a summary and an outlook of follow-up studies. Throughout the main body of this manuscript, we focus on longitudinally polarized virtual photons to simplify our formulas and make our manuscript easier to read. We have performed the same analysis for transversely polarized virtual photons, and the results are shown in appendix~\ref{app:transverse}.

\section{Inclusive back-to-back dihadron cross-section at leading order}

We first consider the LO inclusive dihadron cross-section in small $x$ DIS in the back-to-back limit. We work in the dipole frame where $q^\mu=(q^+,q^-,\boldsymbol{0}_\perp)$ is the four-momentum of the virtual photon with $q^- \gg q^+$ , the polarization of the photon is denoted by $\lambda=\rmL,\rmT$, and $P^\mu=(P^+,M^2/(2P^+),\boldsymbol{0}_\perp)$ is the four momentum of the target with mass $M$. The standard DIS variables are defined as $Q^2=-q^2=-2q^+q^-$ and $x_{\rm Bj}=Q^2/(2P\cdot q)=Q^2/(2P^+q^-)$ when $M\ll P^+$.

\paragraph{Parton-level cross-section.} At leading order in $\alpha_s$, the partonic $\gamma_\lambda^*+A\to q\bar q+X$ cross-section factorizes in the limit $P_\perp\equiv |\Pt|\gg K_\perp\equiv|\Kt|$ where the quark-antiquark pair is produced at forward rapidities (in the direction of the virtual photon) and back-to-back in the transverse plane \cite{Dominguez:2011wm,Dominguez:2011br,Metz:2011wb}:
\begin{align}
    &\left.\frac{\der\sigma^{\gamma_\lambda^\star+A\to q \bar q +X}}{\der^2\ktone\der^2\kttwo\der\eta_1\der\eta_2}\right|_{\rm LO}=\alpha_{\rm em}\alpha_s e_f^2\Hcal_{\rm LO}^{\lambda,ij}(\Pt,Q^2,z_1,z_2)\int\frac{\der^2\bt\der^2\bt'}{(2\pi)^4}e^{-i\Kt\cdot\rbbpt}\hat G^{ij}_\eta(\rbbpt)\,,\label{eq:LO-factorization}
\end{align}
where $\Pt=z_2\ktone-z_1\kttwo$ and $\Kt=\ktone+\kttwo$ are the $q\bar q$ relative transverse momentum and transverse momentum imbalance at the partonic level respectively. We denote the four-momenta of the quark and antiquark as $k_1^\mu$ and $k_2^\mu$, and we define the longitudinal momentum fraction with respect to the virtual photon as $z_i=k_i^-/q^-$. The factorization is written as a product between a hard factor
\begin{align}
    \Hcal_{\rm LO}^{\lambda,ij}=\left\{
    \begin{array}{ll}
        16 \,z_1^3 z_2^3 \,Q^2\, \frac{\Pt^i \Pt^j}{(\Pt^2+z_1z_2 Q^2)^4}  & \mbox{ for }\lambda=\rm L\\
      z_1z_2(z_1^2+z_2^2) \left[\frac{\delta^{ij}}{(\Pt^2+z_1z_2Q^2)^2}-\frac{4z_1z_2Q^2\Pt^i\Pt^j}{(\Pt^2+z_1 z_2Q^2)^4}\right] & \mbox{ for }\lambda=\rm T \,,
    \end{array}
\right. 
\end{align}
and the Weizs\"{a}cker-Williams (WW) gluon TMD distribution (in coordinate space), whose definition in the CGC is
    \begin{align}
    \hat G^{ij}_{\eta}(\rbbpt)&\equiv\frac{-2}{\alpha_s}\left\langle\Tr\left[V(\bt) \left(\partial^iV^\dagger(\bt) \right) V(\bt') \left(\partial^jV^\dagger(\bt') \right)\right]\right\rangle_{\eta}\,,
    \label{eq:WWTMD}
\end{align}
where $V(\bt)$ is a light-like Wilson line in the fundamental representation of $\rm SU(3)$ with transverse coordinate $\bt$. Throughout this paper, we use the shorthand notation 
\begin{align}
    \boldsymbol{r}_{ab}\equiv\at -\bt\,,
\end{align}
for any difference between two transverse coordinates $\at$ and $\bt$. Assuming impact parameter independence of the correlator in Eq.\,\eqref{eq:WWTMD}, the WW gluon TMD only depends on $\rbbpt$. It is customary to decompose the WW gluon TMD in the trace and traceless components:
\begin{align}
    \hat G^{0}_{\eta}(\rbbpt) &= \delta^{ij} \hat G^{ij}_{\eta}(\rbbpt) \,, \\
    \hat h^{0}_{\eta}(\rbbpt) &= \left(\frac{2\rbbpt^i \rbbpt^j}{\rbbpt^2}- \delta^{ij} \right)\hat G^{ij}_{\eta}(\rbbpt) \,.
\end{align}
These are also known as the unpolarized \cite{McLerran:1993ka,McLerran:1993ni,Kovchegov:1996ty} and linearly polarized WW gluon TMDs \cite{Metz:2011wb}. Their momentum-space versions:
\begin{align}
    G^{0}_{\eta}(\Kt) = \int\frac{\der^2\bt\der^2\bt'}{(2\pi)^4}e^{-i\Kt\cdot\rbbpt} \hat G^{0}_{\eta}(\rbbpt) \,, \\
    h^{0}_{\eta}(\Kt) = \int\frac{\der^2\bt\der^2\bt'}{(2\pi)^4}e^{-i\Kt\cdot\rbbpt} \cos(2\Phi) \hat h^{0}_{\eta}(\rbbpt) \,,
\end{align}
where $\Phi$ is the angle between $\Kt$ and $\rbbpt$, satisfy the positivity bound $h^{0}_{\eta}(\Kt) \leq G^{0}_{\eta}(\Kt)$~\cite{Metz:2011wb} in the McLerran-Venugopalan model~\cite{McLerran:1993ka,McLerran:1993ni} and persists after JIMWLK evolution~\cite{Dumitru:2014vka}.

At this perturbative order, the rapidity scale at which one performs the CGC average $\left\langle [...]\right\rangle_\eta$ is arbitrary. We anticipate here that at NLO, one should use the target rapidity variable $\eta=\ln(P^+/k^+)=\ln(1/x)$ whose natural choice $x=x_g$ is such that the plus momentum transferred from the target puts the $q\bar q$ pair on-shell,
\begin{align}
    x_g=\frac{M_{q\bar q}^2+Q^2}{W^2+Q^2}\,,
    \label{eq:xg}
\end{align}
where $M^2_{q\bar q}=\Pt^2/(z_1z_2)$ is the $q\bar q$ invariant mass and $W$ is the $\gamma^*+A$ center-of-mass energy. 
Last, we remind the reader that Eq.\,\eqref{eq:LO-factorization} is only valid up to power corrections in $K_\perp^2/P_\perp^2$ and $Q_s^2/P_\perp^2$ ($Q_s$ implicitly appears in the WW gluon TMD). A detailed study of the relative importance of kinematic (in $K_\perp^2/P_\perp^2$)~\cite{Altinoluk:2019wyu} and ``genuine" saturation\footnote{Genuine saturation corrections refer to contributions to the cross-section that go beyond the WW gluon distribution~\cite{Altinoluk:2019wyu}.} (in $Q_s^2/P_\perp^2$) power corrections for this process can be found in \cite{Mantysaari:2019hkq,Boussarie:2021ybe} (see also \cite{Fujii:2020bkl} for proton-nucleus collisions).

\paragraph{Hadron-level cross-section.} If one measures hadrons in the final state (instead of jets), one must introduce the (bare) fragmentation functions of the quark or the antiquark into hadrons $h_1$ and $h_2$ respectively and labeled $D^0_{h_1/q}(\zeta_1)$, $D^0_{h_2/\bar q}(\zeta_2)$. Here $\zeta_i$ refers to the fraction of the hadron four-momentum $k_{h,i}^{\mu}$ shared by the parton, i.e. $\zeta_i=k_{h,i}^{\mu}/k_i^\mu$. We have then\footnote{To be more precise one should restrict the lower bound in the integration over $\zeta_1$ and $\zeta_2$. The momentum fraction $x_g$ in Eq.\,\eqref{eq:xg} is a function of the partonic kinematics which in turn are functions of $\zeta_1$, $\zeta_2$ and the kinematics of the measured hadrons (see Eq.\,\eqref{eq:parton-hadron-kinamtics}). Requiring $x_g < 1$ constrains the lower bounds of the integration in $\zeta_1$ and $\zeta_2$. }
\begin{align}
    &\left.\frac{\der\sigma^{\gamma_\lambda^\star+A\to h_1 h_2 +X}}{\der^2\ktoneh\der^2\kttwoh\der\eta_{h1}\der\eta_{h2}}\right|_{\rm LO}=\alpha_{\rm em}\alpha_s e_f^2\int_0^1\frac{\der \zeta_1}{\zeta_1^2}\int_0^1\frac{\der \zeta_2}{\zeta_2^2}D^0_{h_1/q}\left(\zeta_1\right)D^0_{h_2/\bar q}\left(\zeta_2\right)\nonumber\\
    &\times\delta(1-z_1-z_2)\Hcal_{\rm LO}^{\lambda,ij}(\Pt,Q^2,z_1,z_2)\int\frac{\der^2\bt\der^2\bt'}{(2\pi)^4}e^{-i\Kt\cdot\rbbpt}\hat G_\eta^{ij}(\rbbpt) \,.
\end{align}
In this expression, $z_1$ and $z_2$ are not independent variables since they are evaluated at $z_i=z_{hi}/\zeta_i$. Recall also that the hard factor is multiplied by a $\delta$-function enforcing $z_1+z_2=1$ or equivalently $z_{h1}/\zeta_1+z_{h2}/\zeta_2=1$. Likewise, the transverse momenta $\Pt$ and $\Kt$ are implicit functions of $\ktoneh$ and $\kttwoh$ since
\begin{align}
    \Pt=\frac{\Pthat}{\zeta_1 \zeta_2}\,,\quad \Kt=\frac{\ktoneh}{\zeta_1}+\frac{\kttwoh}{\zeta_2}=\frac{\boldsymbol{\hat K}_\perp+\frac{\zeta_2-\zeta_1}{\zeta_1 \zeta_2}\Pthat}{z_{h1}+z_{h2}} \,,
    \label{eq:parton-hadron-kinamtics}
\end{align}
where $\Pthat=z_{h2}\ktoneh-z_{h1}\kttwoh$ and $\boldsymbol{\hat K}_\perp=\ktoneh+\kttwoh$ are respectively the \textit{hadronic} relative transverse momentum momentum and transverse momentum imbalance. Since the partonic factorized expression is only valid at leading power in $K_\perp/P_\perp$, it implies one must always ensure that the cross-section is dominated by $\zeta_1$ and $\zeta_2$ configurations where the strong inequality $K_\perp\ll P_\perp$ is fulfilled. This is the case when the \textit{hadrons} are produced back-to-back and if the $\zeta_1$ and $\zeta_2$ integration is dominated by $\zeta_1\sim \zeta_2\sim 1$. However, because of the convolution in $\zeta_1$ and $\zeta_2$, measuring back-to-back hadrons does not necessarily guarantee that the $q\bar q$ pair has been produced in a back-to-back configuration (for instance, if one probes the regime $\zeta_2\sim 1$ and $\zeta_1\ll 1$ which is not kinematically forbidden if $z_{h_1}\ll 1$). We shall therefore restrict ourselves to hadronic final states which imply a back-to-back topology at the partonic level.

\section{Collinear factorization with fragmentation functions for dihadron production in general kinematics}
\label{sec:coll-fact-FF}

We start with a review of the factorization procedure in terms of the collinear fragmentation function to deal with final state collinear divergences. We shall see that although this procedure works when no constraint is imposed on the final state (such as imposing back-to-back kinematics), the NLO impact factor develops large double logarithmic contributions for back-to-back kinematics. However, the coefficient of the double log is not the same as the one obtained for back-to-back \textit{dijet} production. To restore universality, one must go beyond integrated fragmentation functions and consider instead TMD fragmentation functions. This will be done in the last section of this paper.

The proof of the factorization of final state collinear divergences for this process at small-$x$ and in the CGC formalism is not new, see e.g.\,\cite{Bergabo:2022tcu,Iancu:2022gpw} for similar calculations (see also \cite{Fucilla:2023mkl} for the diffractive case). The main difference with respect to these previous approaches is that we implement the factorization of the collinear divergences in terms of fragmentation functions \textit{after} having subtracted the large rapidity logarithms. Although this does not change the collinear evolution of the fragmentation function --- essentially because the plus prescription in the quark DGLAP splitting function naturally regulates the $z_g\to 0$ divergence (with $z_g=k_g^-/q^-$ and $k_g^-$ the longitudinal momentum of the radiated gluon) so that the phase space for final state collinear gluon emissions is well separated from the phase space contributing to B-JIMWLK evolution --- our method enables us to obtain more easily explicit expressions for the remaining finite pieces which build the NLO impact factor. 

Like in Ref.\,\cite{Bergabo:2022tcu}, we use dimensional regularization to isolate the collinear divergence. There are two advantages of doing so: (i) one can properly define the renormalized fragmentation function in the $\overline{\rm{MS}}$ scheme where fragmentation functions are usually provided from fits to experimental data (see e.g.~\cite{deFlorian:2007aj,deFlorian:2007ekg,deFlorian:2014xna}), (ii) it simplifies the combination with the virtual diagrams contributing to the virtual term in the DGLAP equation as such diagrams have also been computed in dimensional regularization in \cite{Caucal:2021ent}.

In this section, we consider the case of a longitudinally polarized photon for simplicity. The calculation for the transversely polarized case is entirely analogous thanks to the factorization of final state collinear gluon emissions.

\subsection{Collinear divergences from real gluon emissions after the shock-wave}
\label{sub:coll-frag}

\begin{figure}
    \centering
    \includegraphics[width=1\textwidth]{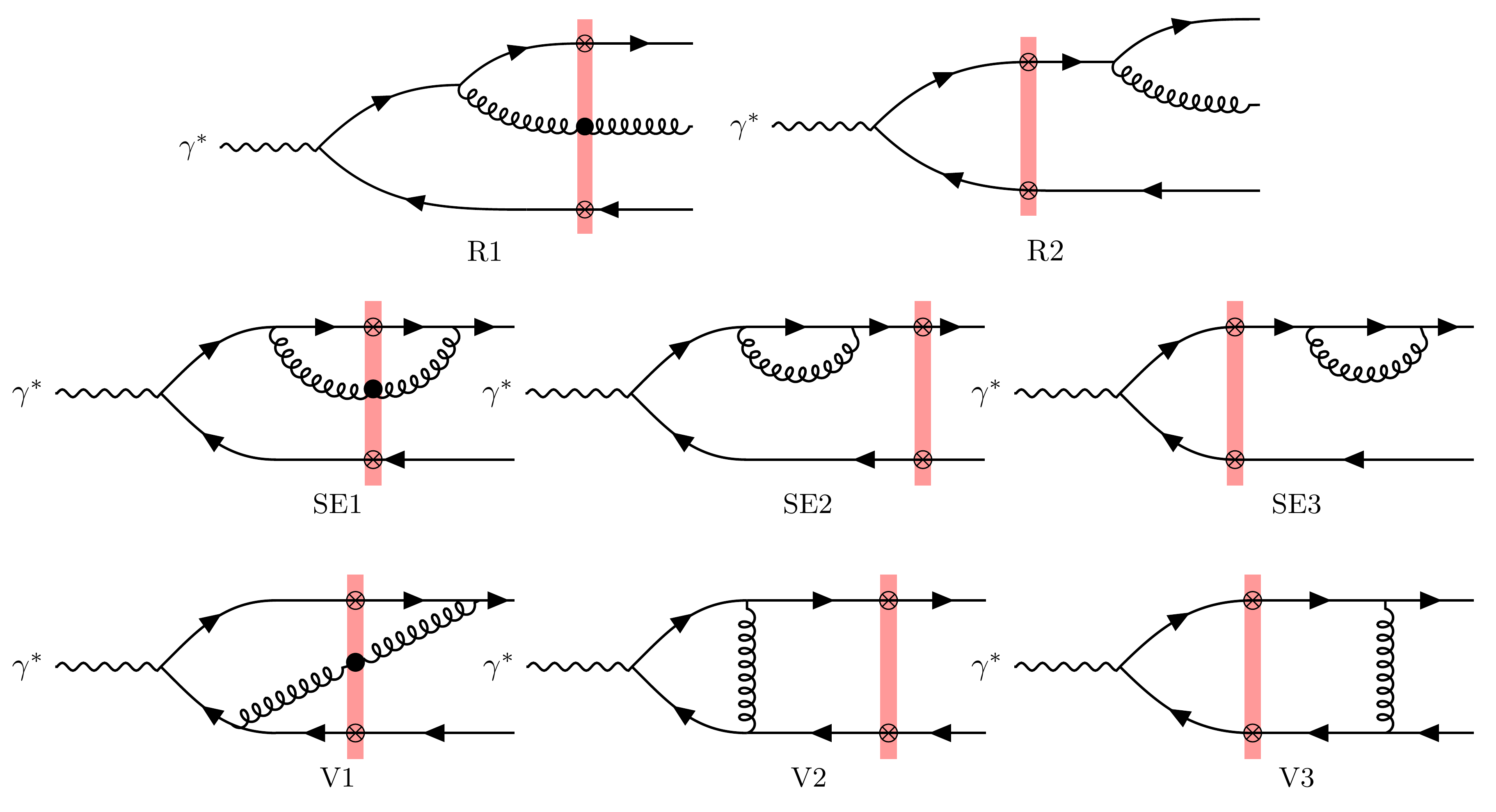}
    \caption{List of NLO Feynman diagrams contributing to the inclusive $q\bar q$ production cross-section in the CGC. Mirror diagrams obtained from $q\leftrightarrow\bar q$ exchange are not shown and are labeled with an additional bar, e.g.\,~\,$\rm R1\to \overline{\rm R1}$. The first, second, and third lines respectively display the real amplitudes, self-energy, and vertex corrections. }
    \label{fig:NLO-graph}
\end{figure}

Our starting point is the real diagram labeled $\rm R2$ where the final state gluon is emitted from the quark. Our conventions for labeling the NLO Feynman graphs contributing to the inclusive dihadron cross-section are explained in Fig.\,\ref{fig:NLO-graph}.
This diagram has an infrared logarithmic singularity when the gluon and the quark are collinear, $\kgt\to z_g/z_1\ktone$. The differential cross-section for inclusive $q\bar q g$ production coming from diagram $\rm R2\times R2^*$ is given by
\begin{align}
      &\left.\frac{\der\sigma^{\gamma_{ \lambda=\rmL}^\star+A\to q \bar qg+X}}{\der^2\ktone\der^2\kttwo\der^2\kgt\der \eta_1\der\eta_2\der\eta_g}\right|_{\rm R2\times R2^*}=\frac{\alpha_{\rm em}e_f^2N_c}{(2\pi)^6}\int_{\xt,\xt',\yt,\yt'} e^{-i\ktone\cdot\rxxtp-i\kttwo\cdot\ryytp}\nonumber\\
    &\times\Xi_{\rm LO}(\xt,\yt;\xt',\yt')\frac{\alpha_sC_F}{\pi^2}8z_1^3z_2^3Q^2\nonumber\\
    &\times\left\{\delta(1-z_1-z_2-z_g)\frac{(1-z_{2})^2}{z_1^2}\left(1+\frac{z_g}{z_1}+\frac{z_g^2}{2z_1^2}\right)K_0(\bar Q_{\mathrm{R}2}r_{xy})K_0(\bar Q_{\mathrm{R}2}r_{x'y'})\frac{e^{-i\kgt \cdot\rxxtp}}{(\kgt-\frac{z_g}{z_1}\ktone)^2}\right.\nonumber\\
    &\left.-\delta(1-z_1-z_2)\Theta(z_f-z_g)K_0(\bar Qr_{xy})K_0(\bar Q r_{x'y'})\frac{e^{-i\kgt \cdot\rxxtp}}{\kgt^2}\right\}\,,
    \label{eq:dijet-NLO-long-R2R2-final}
\end{align}
with $\bar Q_{\rm R2}^2=z_2(1-z_2)Q^2$ and $\bar Q^2=z_1z_2Q^2$. We use the shorthand notation $\int_{\xt} = \int \der^2 \xt$ for the integration over the transverse coordinates $\xt$, etc. The correlator $\Xi_{\rm LO}$ for this particular diagram is identical to the correlator present in the LO cross-section in the CGC, and it reads
\begin{align}
    \Xi_{\rm LO}(\xt,\yt;\xt',\yt')\equiv \left\langle 1-D_{xy}-D_{y'x'}+Q_{xyy'x'}\right\rangle_{z_f} \,,
\end{align}
where 
\begin{align}
  D_{xy}&=\frac{1}{N_c}\Tr[ V(\xt)V^\dagger(\yt)] \,,  \\
  Q_{xyy'x'}&=\frac{1}{N_c}\Tr[ V(\xt)V^\dagger(\yt)V(\yt')V^\dagger(\xt')] \,,
\end{align}
are respectively the dipole correlator and quadrupole correlator. 
The subtraction term in the last line regulates the rapidity divergence as $z_g\to 0$. Only the phase space $z_g\le z_f\equiv k_f^-/q^-$ is subtracted, where $k_f^-$ denotes the rapidity factorization scale along the projectile light-cone direction. In turn, the corresponding counter-term contributes to the small-$x$ or rapidity evolution of the inclusive dihadron cross-section. The notation $\langle [...]\rangle_{z_f}$ in the definition of $\Xi_{\rm LO}$ thus refers to the CGC average over the large $x$ color sources, with a weight functional depending on the rapidity factorization scale $z_f$ according to the B-JIMWLK equation.
The presence of such a subtraction term is a specificity of our calculation at small $x$ as mentioned in the beginning of this section. 

To build the inclusive dihadron cross-section from the partonic $\gamma_{ \lambda}^\star+A\to q \bar qg+X$ cross-section, we introduce the quark and antiquark fragmentation functions and integrate over the gluon four-momentum. As shown in appendix~\ref{app:DGLAP-frag}, the result of this calculation is
\begin{align}
    &\left.\frac{\der\sigma^{\gamma_{ \lambda=\rmL}^\star+A\to h_1h_2+X}}{\der^2\ktoneh\der^2\kttwoh\der \eta_{h1}\der\eta_{h2}}\right|_{\rm R2\times R2^*}=\frac{\alpha_{\rm em}e_f^2N_c}{(2\pi)^6}\frac{(-\alpha_s)}{2\pi}\int_{\xt,\xt',\yt,\yt'}\Xi_{\rm LO}(\xt,\yt;\xt',\yt') \nonumber\\
    &\times\int_0^1\frac{\der \zeta_1}{\zeta_1^2}\int_0^1\frac{\der \zeta_2}{\zeta_2^2}D^0_{h_2/\bar q}(\zeta_2) \ 8z_1^3z_2^3Q^2K_0(\bar Qr_{xy})K_0(\bar Qr_{x'y'})e^{-i\ktone\cdot\rxxtp-i\kttwo\cdot\ryytp}\delta\left(1-z_1-z_2\right)\nonumber\\
    &\times\left\{\int_{\zeta_1}^1\frac{\der \xi}{\xi}P_{qq}(\xi)D^0_{h_1/q}\left(\frac{\zeta_1}{\xi}\right)+C_F\left(2\ln\left(\frac{z_1}{z_f}\right)-\frac{3}{2}\right)D^0_{h_1/q}(\zeta_1)\right\}\left[\frac{2}{\varepsilon}+\ln(\pi\mu^2e^{\gamma_E}\rxxtp^2)\right] \,.\label{eq:final-coll-divergent-real}
\end{align}
At this stage, one should remind the reader that the inclusive $q\bar q$ production cross-section in DIS in the CGC for a longitudinally polarized virtual photon reads
\begin{align}
    \left.\frac{\der\sigma^{\gamma_{ \lambda=\rmL}^\star+A\to q\bar q+X}}{\der^2\ktone\der^2\kttwo\der \eta_{1}\der\eta_{2}}\right|_{\rm LO}&=\frac{\alpha_{\rm em}e_f^2N_c}{(2\pi)^6}\int_{\xt,\xt',\yt,\yt'}e^{-i\ktone\cdot\rxxtp-i\kttwo\cdot\ryytp}\Xi_{\rm LO}(\xt,\yt;\xt',\yt') \nonumber\\
    &\times \delta\left(1-z_1-z_2\right)8z_1^3z_2^3Q^2K_0(\bar Qr_{xy})K_0(\bar Qr_{x'y'})\,,
\end{align}
and that this formula is valid for general $q\bar q$ kinematics at small $x$. Hence, it is clear that the collinear divergence of diagram $\rm R2\times R2^*$ factorizes from the LO cross-section. If one further includes the gluon to hadron fragmentation piece (see appendix~\ref{app:DGLAP-frag}), our final factorized expression reads
\begin{align}
    &\left.\frac{\der\sigma^{\gamma_{ \lambda}^\star+A\to h_1h_2+X}}{\der^2\ktoneh\der^2\kttwoh\der \eta_{h1}\der\eta_{h2}}\right|_{\rm R2\times R2^*}=\int_0^1\frac{\der \zeta_1}{\zeta_1^2}\int_0^1\frac{\der \zeta_2}{\zeta_2^2}D^0_{h_2/\bar q}(\zeta_2)\left.\frac{\der\sigma^{\gamma_{ \lambda}^\star+A\to q\bar q+X}}{\der^2\ktone\der^2\kttwo\der \eta_{1}\der\eta_{2}}\right|_{\rm LO}\nonumber\\
    &\hspace{4cm}\times\frac{(-\alpha_s)}{2\pi} \left\{\int_{\zeta_1}^1\frac{\der \xi}{\xi}P_{qq}(\xi)D^0_{h_1/q}\left(\frac{\zeta_1}{\xi}\right)+\int_{\zeta_1}^1\frac{\der \xi}{\xi}P_{gq}(\xi)D^0_{h_1/g}\left(\frac{\zeta_1}{\xi}\right)\right.\nonumber\\
    &\hspace{4cm}\left.+C_F\left(2\ln\left(\frac{z_1}{z_f}\right)-\frac{3}{2}\right)D^0_{h_1/q}(\zeta_1)\right\}\left[\frac{2}{\varepsilon}+\ln(e^{\gamma_E} \pi\mu^2 \rxxtp^2)\right] \,,
\end{align}
where we have slightly abused the notation since the $\rxxtp$ dependence inside the logarithm should stand inside the transverse coordinate integration in the LO partonic cross-section.

This result is our final result for the calculation of diagram $\rm R2\times R2^*$ in the case of dihadron measurement with quark-antiquark fragmentation. There is of course a similar contribution from diagram $\rm \overline{R2}\times\overline{R2}^*$ obtained by $q\leftrightarrow\bar q$ and $h_1\leftrightarrow h_2$ exchange. In addition to the DGLAP contribution in the first term of the curly bracket, the presence of the term proportional to 
\begin{align}
   C_F \left(2\ln\left(\frac{z_1}{z_f}\right)-\frac{3}{2}\right)D^0_{h_1/q}(\zeta_1) \,, \label{eq:not-DGLAP-term}
\end{align}
may look a little bit odd. Yet, we will see that its presence is crucial to recovering the DGLAP evolution of the fragmentation function and that it will play an important role in the back-to-back limit.

\subsection{Combining the fragmentation contributions with the $1/\varepsilon$ pole from virtual graphs}

Virtual diagrams also display a $1/\varepsilon$ pole which must be combined with the poles from the real cross-section to get the universal collinear evolution of the fragmentation functions. The relevant contribution at the parton level comes from the diagrams $\rm SE2$, $\overline{\rm SE2}$, $\rm V2$ and the singular part of $\rm SE1$ and $\overline{\rm SE1}$~\cite{Caucal:2021ent}; after introducing the fragmentation functions, the sum of these contributions read
\begin{align}
    &\left.\frac{\der\sigma^{\gamma_{ \lambda=\rmL}^\star+A\to q\bar q+X}}{\der^2\ktoneh\der^2\kttwoh\der \eta_{h1}\der\eta_{h2}}\right|_{\rm IR\times LO^*}=\frac{\alpha_{\rm em}e_f^2N_c}{(2\pi)^6}\int_{\xt,\xt',\yt,\yt'} \Xi_{\rm LO}(\xt,\yt;\xt',\yt')\nonumber\\
    & \times \int_0^1\frac{\der \zeta_1}{\zeta_1^2}\int_0^1\frac{\der \zeta_2}{\zeta_2^2}D^0_{h_1/ q}(\zeta_1)D^0_{h_2/\bar q}(\zeta_2)e^{-i\ktone\cdot\rxxtp-i\kttwo\cdot\ryytp}\delta(1-z_1-z_2)\nonumber\\
    &\times 8z_1^3z_2^3Q^2K_0(\bar Qr_{xy})K_0(\bar Qr_{x'y'})\frac{\alpha_sC_F}{2\pi}\left\{\left(\ln\left(\frac{z_1}{z_f}\right)+\ln\left(\frac{z_2}{z_f}\right)-\frac{3}{2}\right)\left(\frac{2}{\varepsilon}+\ln(e^{\gamma_E}\pi\mu^2\rxyt^2)\right)\right.\nonumber\\
    &\left.+\frac{1}{2}\ln^2\left(\frac{z_2}{z_1}\right)-\frac{\pi^2}{6}+\frac{5}{2}\right\} \,.
    \label{eq:dijet-NLO-finite-SE1-V1-SE2uv-hadron}
\end{align}
In this expression, the rapidity divergence has already been subtracted off consistent with the calculation of the previous subsection. This explains the presence of the rapidity factorization scale $z_f$. 
It is now clear that the pole term given by Eq.\,\eqref{eq:not-DGLAP-term} inside Eq.\,\eqref{eq:final-coll-divergent-real} exactly cancels against the pole in Eq.\,\eqref{eq:dijet-NLO-finite-SE1-V1-SE2uv-hadron} (after including the complex conjugate and $q\leftrightarrow\bar q$ exchanged contributions).

In order to properly combine Eq.\,\eqref{eq:dijet-NLO-finite-SE1-V1-SE2uv-hadron} with Eq.\,\eqref{eq:final-coll-divergent-real}, we introduce an arbitrary transverse momentum factorization scale $\mu_F$ and we decompose the logarithms accompanying the $1/\varepsilon$ pole in the real and virtual terms as
\begin{align}
    \ln\left(e^{\gamma_E}\pi\mu^2\rt^2\right)=\ln\left(4\pi e^{-\gamma_E}\right)+\ln\left(\frac{\mu^2}{\mu_F^2}\right)+\ln\left(\frac{\mu_F^2\rt^2}{c_0^2}\right)\,,\label{eq:MSbar-log-dec}
\end{align}
where $\rt$ corresponds to $\rxxtp$ in the real term and $\rxyt$ in the virtual one, and $c_0 = 2 e^{-\gamma_E} $. In the $\overline{\rm MS}$ scheme, the first two logarithms in Eq.\,\eqref{eq:MSbar-log-dec} are removed together with the $1/\varepsilon$ pole. If we define the collinear part of the NLO cross-section as 
\begin{align}
    \left.\frac{\der\sigma^{\gamma_{ \lambda}^\star+A\to h_1h_2+X}}{\der^2\ktoneh\der^2\kttwoh\der \eta_{h1}\der\eta_{h2}}\right|_{\rm coll.}&\equiv\left.\left[\frac{\der\sigma^{\gamma_{ \lambda}^\star+A\to h_1h_2+X}}{\der^2\ktoneh\der^2\kttwoh\der \eta_{h1}\der\eta_{h2}}\right|_{\rm R2\times R2^*}+(h_1\leftrightarrow h_2)\right]\nonumber\\
    &+\left[\left.\frac{\der\sigma^{\gamma_{ \lambda}^\star+A\to h_1h_2+X}}{\der^2\ktoneh\der^2\kttwoh\der \eta_{h1}\der\eta_{h2}}\right|_{\rm IR\times LO^*}+c.c.\right]\,,
\end{align}
we get the following result for this particular combination
\begin{align}
     &\left.\frac{\der\sigma^{\gamma_{ \lambda}^\star+A\to h_1h_2+X}}{\der^2\ktoneh\der^2\kttwoh\der \eta_{h1}\der\eta_{h2}}\right|_{\rm coll.}=\int_0^1\frac{\der \zeta_1}{\zeta_1^2}\int_0^1\frac{\der \zeta_2}{\zeta_2^2}D^0_{h_2/\bar q}(\zeta_2)\left.\frac{\der\sigma^{\gamma_{ \lambda}^\star+A\to q\bar q+X}}{\der^2\ktone\der^2\kttwo\der \eta_{1}\der\eta_{2}}\right|_{\rm LO} \nonumber\\
    &\times \frac{(-\alpha_s)}{2\pi}\left\{\int_{\zeta_1}^1\frac{\der \xi}{\xi}P_{qq}(\xi)D^0_{h_1/q}\left(\frac{\zeta_1}{\xi}\right)+\int_{\zeta_1}^1\frac{\der \xi}{\xi}P_{gq}(\xi)D^0_{h_1/g}\left(\frac{\zeta_1}{\xi}\right)\right\}\left[\frac{2}{\varepsilon}+\ln\left(\frac{4\pi e^{-\gamma_E}\mu^2}{\mu_F^2}\right)\right]\nonumber\\
    &+ \left.\frac{\der\sigma^{\gamma_{ \lambda}^\star+A\to h_1h_2+X}}{\der^2\ktoneh\der^2\kttwoh\der \eta_{h1}\der\eta_{h2}}\right|_{\rm coll.,f}+(h_1\leftrightarrow h_2) \,, \label{eq:combining-real-virtual}
\end{align}
with the finite piece given by 
\begin{align}
    &\left.\frac{\der\sigma^{\gamma_{ \lambda=\rmL}^\star+A\to h_1h_2+X}}{\der^2\ktoneh\der^2\kttwoh\der \eta_{h1}\der\eta_{h2}}\right|_{\rm coll.,f}=\frac{\alpha_{\rm em}e_f^2N_c}{(2\pi)^6}\int_0^1\frac{\der \zeta_1}{\zeta_1^2}\int_0^1\frac{\der \zeta_2}{\zeta_2^2}D^0_{h_2/\bar q}(\zeta_2) \delta\left(1-z_1-z_2\right)\nonumber\\
    &\times\int_{\xt,\xt',\yt,\yt'}\Xi_{\rm LO}(\xt,\yt;\xt',\yt') \ 8z_1^3z_2^3Q^2K_0(\bar Qr_{xy})K_0(\bar Qr_{x'y'})e^{-i\ktone\cdot\rxxtp-i\kttwo\cdot\ryytp}\nonumber\\
    &\times\frac{\alpha_s C_F}{2\pi}\left\{\frac{-1}{C_F}\left[\int_{\zeta_1}^1\frac{\der \xi}{\xi}P_{qq}(\xi)D^0_{h_1/q}(\zeta_1/\xi)+\int_{\zeta_1}^1\frac{\der \xi}{\xi}P_{gq}(\xi)D^0_{h_1/g}(\zeta_1/\xi)\right.\right.\nonumber\\
    &\left.\left.+\left(2\ln\left(\frac{z_1}{z_f}\right)-\frac{3}{2}\right)D^0_{h_1/q}(\zeta_1)\right]\ln\left(\frac{\mu_F^2\rxxtp^2}{c_0^2}\right)+\frac{1}{2}\ln^2\left(\frac{z_2}{z_1}\right)-\frac{\pi^2}{6}+\frac{5}{2}\right.\nonumber\\
    &\left.+D^0_{h_1/q}(\zeta_1)\left[\ln\left(\frac{z_1}{z_f}\right)+\ln\left(\frac{z_2}{z_f}\right)-\frac{3}{2}\right]\ln\left(\frac{\mu_F^2r_{xy}r_{x'y'}}{c_0^2}\right)\right\} \,. \label{eq:combining-real-virtual-1}
\end{align}
In this expression, the first term renormalizes the bare quark fragmentation function in the $\overline{\rm MS}$ scheme ; we can replace in the LO cross-section the bare fragmentation function by the renormalized one at the scale $\mu_F$ such that
\begin{align}
    D_{h_1/q}(\zeta_1,\mu_F^2)&=D^0_{h_1/q}(\zeta_1)-\frac{\alpha_s}{2\pi}\left[\frac{2}{\varepsilon}+\ln\left(4\pi e^{-\gamma_E}\right)+\ln\left(\frac{\mu^2}{\mu_F^2}\right)\right]\nonumber\\
    &\times\left\{\int_{\zeta_1}^1\frac{\der \xi}{\xi}P_{qq}(\xi)D^0_{h_1/q}\left(\frac{\zeta_1}{\xi}\right)+\int_{\zeta_1}^1\frac{\der \xi}{\xi}P_{gq}(\xi)D^0_{h_1/g}\left(\frac{\zeta_1}{\xi}\right)\right\} \,.
\end{align}
The $\mu_F$ independence of the cross-section up to powers of $\alpha_s^2$ corrections yields the standard LO DGLAP equation for the fragmentation function
\begin{align}
    \frac{\partial D_{h_1/q}(\zeta_1,\mu_F^2)}{\partial \ln(\mu_F^2)}&=\frac{\alpha_s}{2\pi}\int_{\zeta_1}^1\frac{\der \xi}{\xi}P_{qq}(\xi)D_{h_1/q}\left(\frac{\zeta_1}{\xi},\mu_F^2\right)+\frac{\alpha_s}{2\pi}\int_{\zeta_1}^1\frac{\der \xi}{\xi}P_{gq}(\xi)D_{h_1/g}\left(\frac{\zeta_1}{\xi},\mu_F^2\right) \,, \label{eq:DGLAP-FF}
\end{align}
where the gluon into hadron piece computed in appendix~\ref{app:DGLAP-frag} has generated flavor mixing~\cite{Bergabo:2024ivx}. To simplify future expressions, we introduce the DGLAP kernel $\mathcal{K}_{\rm DGLAP}$ such that the right hand side of Eq.\,\eqref{eq:DGLAP-FF} is given by $\frac{\alpha_s}{2\pi}\mathcal{K}_{\rm DGLAP}\otimes D_{h_1/q}(\zeta_1,\mu_F)$.

The second term in Eq.\,\eqref{eq:combining-real-virtual-1} are finite pieces that belong to the NLO impact factor for inclusive dihadron production at small-$x$. In the back-to-back limit, they give large Sudakov double logarithmic contributions as we shall discuss in the next section.

In the end, the formula for NLO inclusive dihadron production at small-$x$ reads
\begin{align}
    &\left.\frac{\der\sigma^{\gamma_{ \lambda}^\star+A\to h_1h_2+X}}{\der^2\ktoneh\der^2\kttwoh\der \eta_{h1}\der\eta_{h2}}\right|_{\rm NLO}=\int_0^1\frac{\der \zeta_1}{\zeta_1^2}\int_0^1\frac{\der \zeta_2}{\zeta_2^2}D_{h_1/q}(\zeta_1,\mu_F^2)D_{h_2/q}(\zeta_2,\mu_F^2)\nonumber\\
    &\times \left[\left.\frac{\der\sigma^{\gamma_{ \lambda}^\star+A\to q\bar q+X}}{\der^2\ktone\der^2\kttwo\der \eta_{1}\der\eta_{2}}\right|_{\textrm{LO},k_i^\mu=k_{h,i}^\mu/\zeta_i}+\alpha_s\mathcal{I}(\ktoneh,\kttwoh,z_{h1},z_{h2} ; z_f,\mu_F)\right] \,. \label{eq:NLO-hadron-final}
\end{align}
Implicit in this expression is the small-$x$ resummation: all CGC correlators in the LO cross-section must be evaluated at the rapidity factorization scale $z_f$ according to the B-JIMWLK equation. The NLO impact factor $\mathcal{I}$ has an explicit dependence upon $\mu_F$ and $z_f$ as a result of the factorization of both final-state collinear and rapidity divergences. The explicit calculation of this NLO impact factor in the back-to-back limit is the subject of section~\ref{eq:NLO-if-dihadron}. Before carrying out the systematic computation of the NLO impact factor, we parametrically extract the leading logarithmic terms in the back-to-back limit and discuss the physical meaning of our result.

\section{Heuristic derivation of the Sudakov double logarithms for dihadron}
\label{sec:heuristic-discussion}

In this section, we aim at isolating the leading double logarithmic terms when $P_\perp \gg K_\perp$ in the NLO impact factor for inclusive back-to-back dihadron production and justify from physical grounds why this double logarithm differs between dijet and dihadron measurements. We leave the detailed calculation of the latter in the next section. 

\subsection{Final state corrections with collinear singularity}

We start by estimating the Sudakov double logarithm arising from final state Feynman graphs, without $q\bar q$ interference (those will be discussed in the next subsection). This includes diagrams $\rm R2\times R2^*$, $\overline{\rm R2}\times \overline{\rm R2}^*$, $\rm SE3\times \rm LO^*$ and $\overline{\rm SE3}\times \rm LO^*$ (see Fig.\,\ref{fig:NLO-graph}). One can interpret this Sudakov contribution as a result of a mismatch between real and virtual final state corrections which are collinearly singular. In our regularization scheme\footnote{Dimensional regularization in the transverse momentum and cut-off in the longitudinal momentum.} the final state self-energy corrections vanish for massless quarks, so the virtual corrections that play their role are coming from the combination given in Eq.\,\eqref{eq:dijet-NLO-finite-SE1-V1-SE2uv-hadron} of the $C_F$-dependent diagrams (labeled $\rm SE2$, $\overline{\rm SE2}$, $\rm V2$, and the singular contribution of $\rm SE1$ and $\rm \overline{SE1}$ , see Fig.~\ref{fig:NLO-graph}).  Yet, in a cut-off regularization scheme which is more convenient for our heuristic discussion, the final state self-energies give the counterpart of the final state real correction $\rm R2\times R2^*$ and $\overline{\rm R2}\times\overline{\rm R2}^*$ that cancel the collinear singularity for an infrared and collinear (IRC) safe definition of the cross-section. Real corrections associated with final state soft and collinear gluons contribute as
\begin{align}
    \frac{2\alpha_sC_F}{\pi}\int_{z_f}^1\frac{\der z_g}{z_g}\int_{\Lambda^2}^{\Kt^2}\frac{\der \Lt^2}{ \Lt^2} \,,
\end{align}
where the transverse vector $\Lt=\kgt-z_g/z_1\Pt$ measures the collinearity of the gluon with respect to the quark. Here, the overall factor of $2$ comes from quark and antiquark fragmentation contributions. The boundaries of this double integral are clear from the inspection of Eq.\,\eqref{eq:dijet-NLO-long-R2R2-final}. The $\Lt^2$ integral is cut in the UV by the phase at a scale of order $1/\rbbpt^2\sim \Kt^2$. $\Lambda$ is a cut-off in transverse momentum space, whose dependence disappears in the NLO impact factor by construction. In the longitudinal integral over $z_g$, we exclude gluons contributing to B-JIMWLK evolution, so that $z_g\ge z_f$. This rapidity factorization scale $z_f$ must be chosen of order of $1/(\Pt^2\rbbpt^2)\sim \Kt^2/\Pt^2$~\cite{Caucal:2023fsf}. Physically, this condition comes from imposing gluons with $z_g\le z_f$ contributing to the small-$x$ evolution to have a formation time $\sim z_g q^-/\kgt^2$ smaller than the virtual photon coherence time $\sim q^-/Q^2$~\cite{Beuf:2014uia,Iancu:2015vea,Ducloue:2019ezk}. As $\kgt^2\sim \Kt^2$ for small-$x$ gluons, one gets $z_g\le \kgt^2/Q^2\sim \Kt^2/\Pt^2$ so that $z_f$ must be of order $\Kt^2/\Pt^2$. 

On the other hand, the dominant contributions to the final state self-energies come from the phase space\footnote{According to the Lehmann–Symanzik–Zimmermann (LSZ) reduction formula (see e.g. Sec.\,7.2 in \cite{Peskin:1995ev}), to compute the scattering matrix one should multiply the amputated Green's functions by the wavefunction normalization $Z^{1/2} = 1 + \frac{1}{2} \delta Z$ for each external leg. This amounts to keeping only $\frac{1}{2}$ of the self-energy contributions to the external legs, which explains why the self-energies in Eq.\,\eqref{eq:se-doublelog} come with a factor of 2 instead of 4.}
\begin{align}
    -\frac{2\alpha_sC_F}{\pi}\int_{z_f}^1\frac{\der z_g}{z_g}\int_{\Lambda^2}^{\Pt^2}\frac{\der \Lt^2}{\Lt^2} \,.\label{eq:se-doublelog}
\end{align}
We expect that the $\Pt^2$ scale acts as a UV cut-off, so the phase space $\Lt^2 > \Pt^2$ cancels after including the contribution of all (virtual) diagrams.

The net effect is a Sudakov double logarithm
\begin{align}
    \textrm{DL}_{\rm real-coll.+virtual-pole}&=-\frac{2\alpha_sC_F}{\pi}\int_{z_f}^1\frac{\der z_g}{z_g}\int_{\Kt^2}^{\Pt^2}\frac{\der \Lt^2}{\Lt^2} \,,
    \label{eq:final-state-Sud-heuristic}
\end{align}
giving Eq.\,\eqref{eq:DL-CF-fs} for $z_f\sim \Kt^2/\Pt^2$. The mismatch between the double logarithmic phase space for final state real versus virtual gluon emissions gives a Sudakov double logarithm with coefficient $-2\alpha_s C_F/\pi$.

\subsubsection*{Comparison with Sudakov double log in dijet production}
This should be contrasted with the dijet case, for which the real corrections associated with final state soft and collinear gluons have two contributions, depending on whether the gluon lies inside or outside the quark or antiquark jet. With a cut-off regularization scheme, the in-cone contribution parametrically reads
\begin{align}
    \frac{2\alpha_sC_F}{\pi}\int_{z_f}^{1}\frac{\der z_g}{z_g}\int_{\Lambda^2}^{z_g^2\Pt^2R^2}\frac{\der \Lt^2}{\Lt^2} \,,
\end{align}
where the upper limit comes from the jet cone constraint. This in-cone piece should be combined with the final state self-energies Eq.~\eqref{eq:se-doublelog} which are identical in the dihadron and dijet case, to get
\begin{align}
    -\frac{2\alpha_sC_F}{\pi}\int_{z_f}^1\frac{\der z_g}{z_g}\int_{z_g^2\Pt^2R^2}^{\Pt^2}\frac{\der \Lt^2}{\Lt^2}=-\frac{2\alpha_s C_F}{\pi}\ln^2\left(\frac{\Pt^2}{\Kt^2}\right)-\frac{2\alpha_sC_F}{\pi}\ln^2\left(\frac{\Pt^2}{\Kt^2}\right)\ln\left(\frac{1}{R^2}\right) \,,\label{eq:incone-sudakov}
\end{align}
where we have used $z_f\sim \Kt^2/\Pt^2$ again. The second term in the above expression, which depends on $R$, is a Sudakov single logarithm that we keep for the sake of the comparison between the dijet and dihadron case. In addition to that, a soft and nearly collinear gluon can lie outside the jet cone. When the gluon lies outside the quark jet, it contributes to the NLO cross-section via this double integral, which after subtracting the rapidity divergence, reads
\begin{align}
    &\frac{\alpha_s C_F}{\pi^2}\int_{z_f}^{z_1}\frac{\der z_g}{z_g}\int\der^2\kgt\frac{e^{-i\kgt\cdot\rbbpt}}{\left(\kgt-\frac{z_g}{z_1}\Pt\right)^2}=\frac{\alpha_s C_F}{\pi^2}\int_{z_f}^{z_1}\frac{\der z_g}{z_g}e^{-i\frac{z_g}{z_1}\Pt\cdot\rbbpt}\int\der^2\Lt\frac{e^{-i\Lt\cdot\rbbpt}}{\Lt^2}\,.\label{eq:outcone-calculation}
\end{align}
As far as the boundaries of this double logarithmic integral are concerned, the $z_g$ integral is cut in the infrared by $z_f\sim \Kt^2/\Pt^2$ to exclude gluons accounted for in the small-$x$ evolution. Regarding the $\Lt^2$ integral, it is cut in the UV by the phase at $1/\rbbpt^2\sim \Kt^2$. In the infrared, one must enforce the gluon to be outside the jet; this implies $\Lt^2\ge z_g^2\Pt^2R^2$. This also implies $z_g\le K_\perp/P_\perp$ since $\Lt^2\le \Kt^2$, which is enforced by the phase in the $z_g$ integration in the right-hand side of Eq.\,\eqref{eq:outcone-calculation}. After adding the $(q\leftrightarrow \bar q)$ exchanged diagram, one then gets the following double integral
\begin{align}
\frac{2\alpha_s C_F}{\pi} \int_{K_\perp^2/P_\perp^2}^{K_\perp/P_\perp}\frac{\der z_g}{z_g} \int_{z_g^2\Pt^2R^2}^{\Kt^2}\frac{\der\Lt^2}{\Lt^2}=\frac{2\alpha_s C_F}{4\pi}\ln^2\left(\frac{\Pt^2}{\Kt^2}\right)+\frac{\alpha_sC_F}{\pi}\ln\left(\frac{\Pt^2}{\Kt^2}\right)\ln\left(\frac{1}{R^2}\right) \,, \label{eq:out-of-cone-contrib}
\end{align}
where we also keep track of the $R$ dependent single Sudakov log on the right-hand side.
Combining Eq.\,\eqref{eq:out-of-cone-contrib} and Eq.\,\eqref{eq:incone-sudakov}, we get 
\begin{align}
     \textrm{DL}^{\rm dijet}_{\rm real-coll.+virtual-pole}&=-\frac{3\alpha_s C_F}{2\pi}\ln^2\left(\frac{\Pt^2}{\Kt^2}\right)-\frac{\alpha_sC_F}{\pi}\ln\left(\frac{\Pt^2}{\Kt^2}\right)\ln\left(\frac{1}{R^2}\right)\,, \label{eq:Sudakov-final-state-jet}
\end{align}
where the double logarithm coefficient differs from Eq.\,\eqref{eq:final-state-Sud-heuristic}. To interpret this difference, one observes that in the limit $R\to 0$ where a jet measurement and a hadron measurement should coincide from a physical point of view since the jet is basically made of a single leading hadron, the two results agree with each other. More precisely, for $R^2\sim K_\perp/P_\perp\ll 1$, the $R$ dependent single Sudakov logarithm becomes as large as the double Sudakov logarithm and the two terms in Eq.\,\eqref{eq:Sudakov-final-state-jet} can be combined to reproduce Eq.\,\eqref{eq:final-state-Sud-heuristic}. In this regime, the $\ln(R)$ should also be resummed to all orders, using for instance jet functions~\cite{Dasgupta:2014yra,Kang:2016mcy,Dai:2016hzf,Gutierrez-Reyes:2018qez,Gutierrez-Reyes:2019vbx}.

The difference between the Sudakov double logarithm for inclusive dihadron production and inclusive dijet production thus comes from the phase space for soft and collinear gluon radiations emitted inside the jet cone of opening angle $R$, which does not contribute to the dijet imbalance in the case of a jet measurement. In contrast, it does contribute in the case of a hadronic measurement. As we shall see, this simple physical interpretation makes it natural to associate the additional Sudakov double logarithm seen in inclusive dihadron production to the final state fragmentation process where the recoil of the $q\bar q$ pair against a \textit{very} collinear and soft gluon emission gives a small transverse momentum imbalance to the dihadron system (which could not be measured in the dijet case as such a gluon would lie inside the jets).

\subsection{Final state interferences and virtual gluons crossing the shock-wave}

\label{eq:final-state-inter-and-virtual-cross-SW}

For the other double logarithms, we rely on our calculation of the back-to-back dijet NLO impact factor in \cite{Caucal:2023nci,Caucal:2023fsf}. Indeed, for these contributions, the hadronic and jet cases are completely equivalent since they do not come from the collinearly singular diagrams. Real final state gluons in the interference graphs between the quark and the anti-quark combined with the vertex correction after the shock-wave also give a Sudakov double log equal to 
\begin{align}
    \textrm{DL}_{\rm int}&=-\frac{3\alpha_s}{4\pi N_c}\ln^2\left(\frac{\Pt^2}{\Kt^2}\right) \,.
    \label{eq:interference-Sud-heuristic}
\end{align}
The $1/N_c$ suppression is due to the CGC color structure of diagram $\rm R2\times \overline{\rm R2}^*$, which collapses to the WW operator at leading twist with an overall factor $1/(2N_c)$, which becomes $1/N_c$ after including the $q\leftrightarrow\bar q$ exchange contribution~\cite{Caucal:2022ulg}.
This $-3/4$ coefficient can be easily explained from our discussion of the collinearly divergent diagrams. For the final state interference graphs (either real or virtual), there is no collinear divergence so there is no need to distinguish in-cone vs out-of-cone real gluon contribution. The real gluon emission therefore gives
\begin{align}
\frac{\alpha_s}{\pi N_c}\int_{z_f}^{K_\perp/P_\perp}\frac{\der z_g}{z_g}\int_{z_g^2\Pt^2}^{\Kt^2}\frac{\der\kgt^2}{\kgt^2} \,,
\end{align}
where the bounds of the $\kgt^2$ integral come from the condition of having a double logarithmic phase space, i.e.\,~$1/(\kgt-z_g\Pt)^2\sim 1/\kgt^2$, and the upper bound of the $z_g$ integral ensures that $z_g^2\Pt^2\le \Kt^2$ for consistency. On the other hand, the virtual term yields
\begin{align}
-\frac{\alpha_s}{\pi N_c}\int_{z_f}^{1}\frac{\der z_g}{z_g}\int_{z_g^2\Pt^2}^{\Pt^2}\frac{\der\kgt^2}{\kgt^2} \,,
\end{align}
since for virtual corrections, the upper bound for the gluon transverse momentum is the hard scale $\Pt$ (in turn, the upper limit for $z_g$ simply becomes 1 since $z_g^2\Pt^2$ is smaller than $\Pt^2$ provided $z_g\le 1$).
Combining real and virtual gives the overall factor $-3\alpha_s/(4\pi N_c)$ in front of the double log $\ln^2(\Pt^2/\Kt^2)$.

Finally, the combination of the dressed self-energy diagrams (graphs $\rm SE1$ and $\overline{\rm SE1}$) and vertex corrections (graphs $\rm V1$ and $\overline{\rm V1}$) where the gluon crosses the shock-wave contribute also to the Sudakov double logarithm~\cite{Caucal:2023nci,Caucal:2023fsf}. In these diagrams, the gluon is emitted before the shock-wave\footnote{Initial state self-energy and vertex corrections (diagrams $\rm SE2$, $\overline{\rm SE2}$ and $\rm V2$), where the gluon is also emitted before the scattering, do not contribute to the Sudakov double logarithm, see also \cite{Mueller:2013wwa}.}; their dynamics are therefore sensitive to the decay of the virtual photon into the $q\bar q$ pair. More specifically, the NLO light-cone wave function of the virtual photon into a $q\bar q$ pair plus a virtual gluon is exponentially suppressed when $\kgt^2\ll z_g\Pt^2$. After combining these diagrams, one gets the following double logarithmic phase-space, 
\begin{align}
    \textrm{DL}_{\rm cross}&=
    \frac{\alpha_s N_c}{\pi} \int_{z_f}^1\frac{\der z_g}{z_g}\int_{z_g\Pt^2}^{\Pt^2} \frac{\der \kgt^2}{\kgt^2}\,, \\
    & =\frac{\alpha_s N_c}{2\pi}\ln^2\left(\frac{\Pt^2}{\Kt^2}\right) \,.
    \label{eq:cross-Sud-heuristic}
\end{align}
Here, the overall factor of $N_c$ comes again from the leading twist expansion of the CGC color structure of $\rm SE1$, $\overline{\rm SE1}$, $\rm V1$ $\overline{\rm V1}$ which contains both $C_F$ and $N_c$ dependent terms (see Eq.\,(3.30) in~\cite{Caucal:2023fsf}). To double logarithmic accuracy, the $C_F$ dependent term cancels between the self-energy $\rm SE1$ and vertex $\rm V1$ diagrams (and likewise for the $q\leftrightarrow\bar q$ exchange diagrams), so that only the $N_c$ contribution survives once combined. This is to be expected since the B-JIMWLK evolution of the WW gluon TMD has an overall $N_c$ factor~\cite{Dominguez:2011gc} and since this contribution is tight to the separation between Sudakov and small-$x$ dynamics. In the end, the role of this contribution is indeed to effectively enforce the $\kgt$-dependent lifetime ordering constraint $z_g\ge \kgt^2/Q^2\sim \kgt^2/\Pt^2$ for gluons contributing to the Sudakov double logarithm (and not to B-JIMWLK evolution), instead of the weaker constraint $z_g\ge z_f\sim \Kt^2/\Pt^2$.

\subsection{Discussion}

Combining all Sudakov double logs (Eqs.\eqref{eq:final-state-Sud-heuristic},\,\eqref{eq:interference-Sud-heuristic} and \eqref{eq:cross-Sud-heuristic}), we get
\begin{align}
    \textrm{DL}&\equiv \textrm{DL}_{\rm real-coll.+virtual-pole}+\textrm{DL}_{\rm cross}+\textrm{DL}_{\rm int} \nonumber \\
    &=-\frac{\alpha_s}{\pi}\left[\frac{C_F}{4}+\frac{C_F}{4}+\frac{N_c}{4}\right]\ln^2\left(\frac{\Pt^2}{\Kt^2}\right) \,. \label{eq:double-log-dihadron}
\end{align}
The total Sudakov double logarithm for inclusive dihadron production is not identical to the Sudakov double logarithm for inclusive dijet production, which has a coefficient $-\alpha_s N_c/(4\pi)$. The difference is a factor $-\alpha_s C_F/(4\pi)$ for each outgoing parton in the final state (that is why we display Eq.\,\eqref{eq:double-log-dihadron} in this way), and it entirely comes from the way final state collinear divergences are regulated, via collinear fragmentation functions or a jet definition.

Let us summarize the main message of this section. After the renormalization of the quark/antiquark into hadron fragmentation functions to take care of the final state collinear divergences and the renormalization of the CGC correlators to take care of the large small-$x$ logarithms, the finite pieces in the NLO cross-section for inclusive dihadron production develops large double logarithmic contributions in the ratio $P_\perp/K_\perp \gg 1$. The coefficient of the double log is given by Eq.\,\eqref{eq:double-log-dihadron} and differs from the Sudakov double log for inclusive \textit{dijet} production in DIS at small $x$. Physically, this difference can be explained by noticing that soft-collinear gluons emitted inside the jet cone in the case of a dijet measurement do not contribute to the dijet imbalance, although they would contribute to the dihadron imbalance if one were measuring the imbalance between the two leading hadrons that build the jets. 

We also observed that in the limit $R \to 0$ or more precisely $R^2 \sim K_\perp/P_\perp \ll 1$, the $R$ dependent single Sudakov logarithm becomes as large as the Sudakov double log and the dihadron and dijet Sudakov logs become of comparable magnitude. This is expected as the jet measurement for a very narrow cone reduces to a single leading hadron measurement.

\section{NLO impact factor in the back-to-back limit}
\label{eq:NLO-if-dihadron}

In this section, we compute the back-to-back (in the transverse plane) limit of the NLO impact factor for inclusive dihadron production in DIS at small-$x$, including the dominant Sudakov double and single logarithms, as well as the $\mathcal{O}(\alpha_s)$ corrections which are not enhanced by any logarithm of $1/x$ (B-JIMWLK), $K_\perp/\Lambda_{\rm QCD}$ (DGLAP) or $P_\perp/K_\perp$ (Sudakov). As in the LO case, we keep only the leading power contributions in $\Kt^2/\Pt^2$ and $Q_s^2/\Pt^2$, but all orders in $Q_s^2/\Kt^2$. For diagrams giving the same contributions between dihadron and dijet production (in the small jet radius limit), we rely on the results presented in \cite{Caucal:2023nci,Caucal:2023fsf}. We emphasize that our extraction of the leading power contribution is subject to the same caveats as in the LO case: we assume that the hadronic kinematics are such that it implies a back-to-back \textit{partonic} configuration. 

\subsection{NLO impact factor from the finite pieces after final state collinear factorization}
\label{sub:NLOcoef-coll-fact}

We start by considering the finite pieces left over after ``absorbing" the collinear divergence into fragmentation functions. Our starting point is therefore Eq.\,\eqref{eq:combining-real-virtual-1} where 
the bare fragmentation functions are replaced by the renormalized ones
\begin{align}
    &\left.\frac{\der\sigma^{\gamma_{ \lambda=\rmL}^\star+A\to h_1h_2+X}}{\der^2\ktoneh\der^2\kttwoh\der \eta_{h1}\der\eta_{h2}}\right|_{\rm coll,f}
    =\frac{\alpha_{\rm em}e_f^2N_c}{(2\pi)^6} \int_0^1\frac{\der \zeta_1}{\zeta_1^2}\int_0^1\frac{\der \zeta_2}{\zeta_2^2}\int_{\xt,\xt',\yt,\yt'}\Xi_{\rm LO}(\xt,\yt;\xt',\yt') \nonumber\\
    &\times D_{h_2/\bar q}(\zeta_2,\mu_F^2) \ 8z_1^3z_2^3Q^2K_0(\bar Qr_{xy})K_0(\bar Qr_{x'y'})e^{-i\ktone\cdot\rxxtp-i\kttwo\cdot\ryytp}\delta\left(1-z_1-z_2\right)\nonumber\\
    &\times\frac{\alpha_sC_F}{2\pi}\left\{-\left[\frac{1}{C_F}\mathcal{K}_{\rm DGLAP}\otimes D_{h_1/q}(\zeta_1,\mu_F^2)+\left(2\ln\left(\frac{z_1}{z_f}\right)-\frac{3}{2}\right)D_{h_1/q}(\zeta_1,\mu_F^2)\right]\ln\left(\frac{\mu_F^2\rxxtp^2}{c_0^2}\right)\right.\nonumber\\
    &\left.+D_{h_1/q}(\zeta_1,\mu_F^2)\left[\ln\left(\frac{z_1z_2}{z_f^2}\right)-\frac{3}{2}\right]\ln\left(\frac{\mu_F^2r_{xy}r_{x'y'}}{c_0^2}\right)+\frac{1}{2}\ln^2\left(\frac{z_2}{z_1}\right)-\frac{\pi^2}{6}+\frac{5}{2}\right\}+(h_1\leftrightarrow h_2) \,. \label{eq:combining-real-virtual-2}
\end{align}
We then change the variables from $\xt$, $\yt$ to $\ut=\xt-\yt$ and $\bt=z_1\xt+z_2\yt$ respectively conjugated to $\Pt$ and $\Kt$. The color correlator $\Xi_{\rm LO}$ can be expanded to the first non-trivial order for small $u_\perp \ll b_\perp$~\cite{Dominguez:2011wm},
\begin{equation}
    \Xi_{\mathrm{LO}}(\xt,\yt;\xt',\yt')\approx\ut^i\ut'^j\times\frac{(-\alpha_s)}{2N_c}\hat G^{ij}_{z_f}(\rbbpt) \,,
    \label{eq:correlation_limit_expansion}
\end{equation}
and the resulting transverse coordinate integral over $\ut$ and $\ut'$ are performed analytically such that in the back limit, the above expression becomes
\begin{align}
      &\left.\frac{\der\sigma^{\gamma_{ \lambda=\rmL}^\star+A\to h_1h_2+X}}{\der^2\ktoneh\der^2\kttwoh\der \eta_{h1}\der\eta_{h2}}\right|_{\rm coll.,f}
    =\alpha_s\alpha_{\rm em}e_f^2\int_0^1\frac{\der \zeta_1}{\zeta_1^2}\int_0^1\frac{\der \zeta_2}{\zeta_2^2}D_{h_2/\bar q}(\zeta_2,\mu_F^2)D_{h_1/q}(\zeta_1,\mu_F^2) \ \nonumber\\
    &\times \ \delta\left(1-z_1-z_2\right)\Hcal_{\rm LO}^{\lambda=\rmL,ij}(\Pt,Q,z_1,z_2)\int\frac{\der^2\bt\der^2\bt'}{(2\pi)^4}  e^{-i\Kt\cdot\rbbpt} \hat G^{ij}_{z_f}(\rbbpt)\nonumber\\
    &\times\frac{\alpha_sC_F}{2\pi}\left\{-\left[\frac{\mathcal{K}_{\rm DGLAP}\otimes D_{h_1/q}(\zeta_1,\mu_F^2)}{C_F D_{h_1/q}(\zeta_1,\mu_F^2)}+\left(2\ln\left(\frac{z_1}{z_f}\right)-\frac{3}{2}\right)\right]\ln\left(\frac{\mu_F^2\rbbpt^2}{c_0^2}\right)\right.\nonumber\\
    &\left.-\left[\ln\left(\frac{z_1z_2}{z_f^2}\right)-\frac{3}{2}\right]\left[\ln\left(\frac{c_0^2\Pt^2}{\mu_F^2}\right)-\frac{\Hcal_{\rm NLO,1}^{\lambda=\rmL,ij}}{2\Hcal_{\rm LO}^{\lambda=\rmL,ij}}\right]+\frac{1}{2}\ln^2\left(\frac{z_2}{z_1}\right)-\frac{\pi^2}{6}+\frac{5}{2}\right\}+(h_1\leftrightarrow h_2)\,.\label{eq:combining-real-virtual-3}
\end{align}
The $\rm NLO,1$ hard function for a longitudinally polarized virtual photon is given by
\begin{align}
    \Hcal_{\rm NLO,1}^{\lambda=\rmL,ij}&\equiv4 z_1^3z_2^3Q^2\int\frac{\der^2\ut\der^2\ut'}{(2\pi)^2}e^{-i\Pt\cdot\ruupt}\ut^i\ut'^jK_0(\bar Q u_\perp)K_0(\bar Qu_\perp')\ln(\Pt^4\ut^2\ut'^2)\,, \nonumber \\
    &=\Hcal_{\rm LO}^{\lambda=\rmL,ij}\times \left[4-4\ln\left(\frac{\chi}{c_0}\right)-4\ln\left(1+\frac{1}{\chi^2}\right)\right]\,, \label{eq:hard-NLO1}
\end{align}
and the variable $\chi$ is defined as
\begin{align}
    \chi\equiv \frac{Q}{M_{q\bar q}}=\frac{\sqrt{z_1z_2}Q}{P_\perp}\,.
\end{align}
It comes from the term proportional to $\ln(\mu_F^2r_{xy}r_{x'y'}/c_0^2)$ in Eq.\,\eqref{eq:combining-real-virtual-2} where the logarithm is split into two terms $\ln(\mu_F^2/(c_0^2\Pt^2))+\ln(\Pt^2r_{xy}r_{x'y'})$. The integral in Eq.\,\eqref{eq:hard-NLO1} has been carried out in Sec.\,(3.2) in \cite{Caucal:2022ulg}.

Let us investigate the leading behavior of this particular contribution to the NLO impact factor for $P_\perp\gg K_\perp$. In the back-to-back limit and to leading double logarithmic accuracy, we can replace $|\rxxtp|$ by $|\rbbpt|\sim 1/K_\perp$ and $r_{xy}\sim r_{x'y'}\sim 1/P_\perp$ up to power corrections in $\Kt^2/\Pt^2$ or $Q_s^2/\Pt^2$. As discussed earlier, the rapidity factorization scale $z_f$ must be chosen $\mathcal{O}(\Kt^2/\Pt^2)$.
Using these parametric estimates for the several scales entering into Eq.\,\eqref{eq:combining-real-virtual-3}, the leading term is therefore
\begin{align}
   & \frac{(-\alpha_s)C_F}{2\pi}\left[\int_{\zeta_1}^1\frac{\der \xi}{\xi}P_{qq}(\xi)D^0_{h_1/q}(\zeta_1/\xi)+\left(2\ln\left(\frac{z_1\Pt^2}{\Kt^2}\right)-\frac{3}{2}\right)D^0_{h_1/q}(\zeta_1)\right]\ln\left(\frac{\mu_F^2}{\Kt^2}\right)\nonumber\\
    &+\frac{\alpha_sC_F}{2\pi}D^0_{h_1/q}(\zeta_1)\left[\ln\left(\frac{z_1\Pt^2}{\Kt^2}\right)+\ln\left(\frac{z_2\Pt^2}{\Kt^2}\right)-\frac{3}{2}\right]\ln\left(\frac{\mu_F^2}{\Pt^2}\right)+(h_1\leftrightarrow h_2) \,.
\end{align}
A natural choice of $\mu_F$ would be $\sim K_\perp$ in the first term (the real one) as it minimizes the magnitude of this finite piece. But then, the second term (the virtual one) develops a large logarithm $\ln(\Pt^2/\Kt^2)$ in the back-to-back limit. Vice-versa, if one chooses $\mu_F\sim P_\perp$ to minimize the virtual term, the first term becomes large and the double log extracted in this way is the same as for $\mu_F\sim K_\perp$. Let us then stick to our first choice for the factorization scale $\mu_F\sim 1/r_{bb'}\sim K_\perp$, which presents the advantage of simplifying the form of the single Sudakov logarithm.

Using these physically motivated estimates for the factorization scale, the collinearly divergent real diagrams combined with the virtual graphs with a $1/\varepsilon$ pole then yields the following Sudakov double logarithm
\begin{align}
    \textrm{DL}_{\rm real-coll.+virtual-pole}&=-\frac{2\alpha_s C_F}{\pi}\ln^2\left(\frac{\Pt^2}{\Kt^2}\right) \,, \label{eq:DL-CF-fs}
\end{align}
in agreement with our heuristic discussion in the previous section. Eq.\,\eqref{eq:combining-real-virtual-3} is then a genuine NLO correction without large small-$x$ nor collinear logarithms, provided one chooses $\mu_F\sim c_0/r_{bb'}$ to minimize the residual DGLAP-like collinear logarithm in the third line. Yet, it contains large Sudakov logs and in particular the double logarithm Eq.\,\eqref{eq:DL-CF-fs}.

\subsection{NLO impact factor from final state interferences}

We consider now the real corrections where a final state gluon is emitted by the quark and absorbed by the antiquark in the complex conjugate amplitude (and vice versa) and the final state vertex corrections. These diagrams have been computed in the dijet case and do not have a collinear singularity. Therefore, the result in the dihadron case is identical to the dijet case, modulo the convolution with the quark or antiquark into hadron fragmentation functions. An interesting feature of this contribution is that the real final state gluons emitted at large angles induce anisotropy in the angle between $\Pt$ and $\Kt$ (at the partonic level)~\cite{Hatta:2020bgy,Hatta:2021jcd,Caucal:2022ulg}. We thus need to decompose the parton-level cross-section into Fourier harmonics according to 
\begin{align}
        \frac{\der \sigma^{\gamma_{\lambda}^{\star}+A\to q\bar q+X}}{ \der^2 \Pt \der^2 \Kt \der \eta_1 \der \eta_{2}}= \der \sigma^{(0),\lambda}(P_\perp,K_\perp,\eta_1,\eta_2)+2\sum_{n=1}^{\infty} \der \sigma^{(n),\lambda}(P_\perp,K_\perp,\eta_1,\eta_2)\cos(n\phi)\,,\label{eq:Fourier-dec}
\end{align}
where $\phi$ is the azimuthal angle between $\Pt$ and $\Kt$. Only the even harmonics contribute due to charge-parity symmetry. Each Fourier component is then convoluted with the quark and antiquark fragmentation functions into hadrons; hence we decompose the differential cross-section at the hadronic level as
\begin{align}
    \frac{\der\sigma^{\gamma_{ \lambda}^\star+A\to h_1h_2+X}}{\der^2\ktoneh\der^2\kttwoh\der \eta_{h1}\der\eta_{h2}} = \sum_{n=0}^\infty \frac{\der\sigma^{\gamma_{ \lambda}^\star+A\to h_1h_2+X, (n)}}{\der^2\ktoneh\der^2\kttwoh\der \eta_{h1}\der\eta_{h2}} \,,\label{eq:hadron-Fourier-def}
\end{align}
where 
\begin{align}
    \frac{\der\sigma^{\gamma_{ \lambda}^\star+A\to h_1h_2+X, (n)}}{\der^2\ktoneh\der^2\kttwoh\der \eta_{h1}\der\eta_{h2}}\equiv \int_0^1\frac{\der \zeta_1}{\zeta_1^2}\int_0^1\frac{\der \zeta_2}{\zeta_2^2}D_{h_2/\bar q}(\zeta_2,\mu_F^2)D_{h_1/q}(\zeta_1,\mu_F^2)  \kappa_{n} \der\sigma^{(n),\lambda} \cos(n \phi)\,,\label{eq:hadron-Fourier-dec}
\end{align}
where $\kappa_0 = 1$ and $\kappa_{n} = 2$ for $n >0$. 

At LO, we remind the reader that only the zeroth and second parton-level Fourier modes are non-vanishing, such that 
\begin{align}
    \left.\der \sigma^{(0),\lambda=\rm L}\right|_{\rm LO}&=\alpha_s\alpha_{\rm em}e_f^2\delta(1-z_1-z_2)\Hcal_{\rm LO}^{\lambda=\rmL,0}\int\frac{\der^2\bt\der^2\bt'}{(2\pi)^4}  e^{-i\Kt\cdot\rbbpt}\hat G^{0}_{z_f}(\rbbpt) \,, \\
    \left.\der \sigma^{(2),\lambda=\rm L}\right|_{\rm LO}&=\alpha_s\alpha_{\rm em}e_f^2\delta(1-z_1-z_2)\Hcal_{\rm LO}^{\lambda=\rmL,0}\int\frac{\der^2\bt\der^2\bt'}{(2\pi)^4}  e^{-i\Kt\cdot\rbbpt}\frac{\cos\left(2 \Phi\right)}{2}\hat h^{0}_{z_f}(\rbbpt) \,,
\end{align}
where $\hat G^{0}_{z_f}(\rbbpt)$ and $\hat h^{0}_{z_f}(\rbbpt)$ are the unpolarized and linearly polarized WW gluon TMDs respectively, and $\Phi$ is the azimuthal angle between $\Kt$ and $\rbbpt$. We note that because of the convolution with fragmentation functions, the Fourier modes in the Fourier decomposition of the hadron level cross-section --- as a function of the angle $\phi_h$ between $\Pthat$ and $\boldsymbol{\hat K}$ --- are mixtures of the parton-level Fourier harmonics in Eq.\,\eqref{eq:Fourier-dec}. For this reason, we also provide detailed expressions for all these parton-level Fourier modes.

From the results obtained in \cite{Caucal:2023nci,Caucal:2023fsf}, the contribution coming from the $n=0$ Fourier component reads
\begin{align}
      &\left.\frac{\der\sigma^{\gamma_{ \lambda=\rmL}^\star+A\to h_1h_2+X, (0)}}{\der^2\ktoneh\der^2\kttwoh\der \eta_{h1}\der\eta_{h2}}\right|_{\rm int}
    =\alpha_s\alpha_{\rm em}e_f^2\int_0^1\frac{\der \zeta_1}{\zeta_1^2}\int_0^1\frac{\der \zeta_2}{\zeta_2^2}D_{h_2/\bar q}(\zeta_2,\mu_F^2)D_{h_1/q}(\zeta_1,\mu_F^2) \ \nonumber\\
    &\times \ \delta\left(1-z_1-z_2\right)\Hcal_{\rm LO}^{\lambda=\rmL,0}(\Pt,Q,z_1,z_2)\int\frac{\der^2\bt\der^2\bt'}{(2\pi)^4}  e^{-i\Kt\cdot\rbbpt}\hat G^{0}_{z_f}(\rbbpt)\nonumber\\
    &\frac{\alpha_s}{2\pi N_c}\left\{\frac{1}{2}\ln^2\left(\frac{\Pt^2\rbbpt^2}{c_0^2}\right)-\ln(z_1z_2)\ln\left(\frac{\Pt^2\rbbpt^2}{c_0^2}\right)-\frac{\Hcal_{\rm NLO,2}^{\lambda=\rmL,ii}}{2\Hcal_{\rm LO}^{\lambda=\rmL,0}}\right.\nonumber\\
    &\left.-\ln\left(\frac{z_1z_2}{z_f^2}\right)\left[\ln\left(\frac{\Pt^2\rbbpt^2}{c_0^2}\right)+\ln(c_0^2)-\frac{\Hcal_{\rm NLO,1}^{\lambda=\rmL,ii}}{4\Hcal_{\rm LO}^{\lambda=\rmL,0}}\right]\right\}\nonumber\\
    &+\alpha_s\alpha_{\rm em}e_f^2\int_0^1\frac{\der \zeta_1}{\zeta_1^2}\int_0^1\frac{\der \zeta_2}{\zeta_2^2}D_{h_2/\bar q}(\zeta_2,\mu_F^2)D_{h_1/q}(\zeta_1,\mu_F^2) \ \nonumber\\
    &\times \ \delta\left(1-z_1-z_2\right)\Hcal_{\rm LO}^{\lambda=\rmL,0}(\Pt,Q,z_1,z_2)\int\frac{\der^2\bt\der^2\bt'}{(2\pi)^4}  e^{-i\Kt\cdot\rbbpt}\hat h^{0}_{z_f}(\rbbpt)\frac{\alpha_s}{2\pi N_c}\left[1-\ln(z_1z_2)\right] \,.\label{eq:interf+v3}
\end{align}
Using that $z_f\sim K_\perp^2/P_\perp^2$, one checks that the Sudakov double logarithm coming from final state interferences has a coefficient $-3\alpha_s/(4\pi N_c)$ as parametrically illustrated in Sec.\,\ref{eq:final-state-inter-and-virtual-cross-SW}.
The $\rm NLO,2$ hard function coming from the final state vertex correction reads\footnote{We note that the expression provided in \cite{Caucal:2023nci} contains typos for the hard coefficient functions $\mathcal{H}^{\lambda=\rmL,ij}_{\rm NLO,2}$ and $\mathcal{H}^{\lambda=\rmL,ij}_{\rm NLO,4}$, which nevertheless do not alter the final result presented in \cite{Caucal:2023nci}.}
\begin{align}
    &\mathcal{H}^{\lambda=\rmL,ij}_{\rm NLO,2}(\Pt,Q,z_1,z_2)=\mathcal{H}^{\lambda=\rmL,ij}_{\rm LO}(\Pt,Q,z_1,z_2)\left[\frac{\pi^2}{12}-\ln(\chi)-\frac{1}{2}\ln(z_2)\ln\left(\frac{z_1}{z_2}\right)\right.\nonumber\\
    &+\frac{1}{2}\textrm{Li}_2\left(\frac{z_2-z_1\chi^2}{z_2}\right)+\frac{1}{2}\textrm{Li}_2\left(-\frac{z_1}{z_2}\right)-\frac{1}{2}\textrm{Li}_2\left(z_2-z_1\chi^2\right)-\frac{1}{2}\textrm{Li}_2\left(\frac{z_1\chi^2-z_2}{\chi^2}\right)\nonumber\\
    &+\frac{(z_1-3z_2)z_2+(1+8z_1z_2)\chi^2+z_1(z_2-3z_1)\chi^4}{2(z_2-z_1\chi^2)^2}\ln\left(\frac{z_2(1+\chi^2)}{\chi^2}\right)\nonumber\\
    &\left.-2\textrm{Li}_2\left(\frac{z_2-z_1\chi^2}{z_2(1+\chi^2)}\right)-\frac{z_1(1+\chi^2)}{z_2-z_1\chi^2}+\left(1\leftrightarrow 2\right)\right] \,,
\end{align}
and it depends on the dilogarithm function defined as
\begin{align}
    \textrm{Li}_2(x)\equiv -\int_0^x\frac{\ln(1-u)}{u}\der u\,, 
\end{align}
for $x<1$. Likewise, the second Fourier harmonic ($n=2$) arising from final state interferences is given by
\begin{align}
      &\left.\frac{\der\sigma^{\gamma_{ \lambda=\rmL}^\star+A\to h_1h_2+X, (2)}}{\der^2\ktoneh\der^2\kttwoh\der \eta_{h1}\der\eta_{h2}}\right|_{\rm int}
    =\alpha_s\alpha_{\rm em}e_f^2\int_0^1\frac{\der \zeta_1}{\zeta_1^2}\int_0^1\frac{\der \zeta_2}{\zeta_2^2}D_{h_2/\bar q}(\zeta_2,\mu_F^2)D_{h_1/q}(\zeta_1,\mu_F^2) \ \nonumber\\
    &\times \ \cos(2\phi)\delta\left(1-z_1-z_2\right)\Hcal_{\rm LO}^{\lambda=\rmL,0}(\Pt,Q,z_1,z_2)\int\frac{\der^2\bt\der^2\bt'}{(2\pi)^4}  e^{-i\Kt\cdot\rbbpt}\cos(2\Phi)\hat h^{0}_{z_f}(\rbbpt)\nonumber\\
    &\frac{\alpha_s}{2\pi N_c}\left\{\frac{1}{2}\ln^2\left(\frac{\Pt^2\rbbpt^2}{c_0^2}\right)-\ln(z_1z_2)\ln\left(\frac{\Pt^2\rbbpt^2}{c_0^2}\right)-\frac{5}{4}+\frac{1}{2}\ln(z_1z_2)-\frac{\Hcal_{\rm NLO,2}^{\lambda=\rmL,ii}}{2\Hcal_{\rm LO}^{\lambda=\rmL,0}}\right.\nonumber\\
    &\left.-\ln\left(\frac{z_1z_2}{z_f^2}\right)\left[\ln\left(\frac{\Pt^2\rbbpt^2}{c_0^2}\right)+\ln(c_0^2)-\frac{\Hcal_{\rm NLO,1}^{\lambda=\rmL,ii}}{4\Hcal_{\rm LO}^{\lambda=\rmL,0}}\right]\right\}\nonumber\\
    &+\alpha_s\alpha_{\rm em}e_f^2\int_0^1\frac{\der \zeta_1}{\zeta_1^2}\int_0^1\frac{\der \zeta_2}{\zeta_2^2}D_{h_2/\bar q}(\zeta_2,\mu_F^2)D_{h_1/q}(\zeta_1,\mu_F^2)\cos(2\phi) \ \nonumber\\
    &\times \ \delta\left(1-z_1-z_2\right)\Hcal_{\rm LO}^{\lambda=\rmL,0}(\Pt,Q,z_1,z_2)\int\frac{\der^2\bt\der^2\bt'}{(2\pi)^4}  e^{-i\Kt\cdot\rbbpt}\cos\left(2 \Phi\right)\hat G^{0}_{z_f}(\rbbpt)\nonumber\\
    &\times \frac{\alpha_s}{\pi N_c}\left[1-\ln(z_1z_2)\right] \,.\label{eq:interf+v3-n=2}
\end{align}
For the higher order Fourier components $n>2$, we refer the reader to the expressions displayed in the summary section~\ref{sub:sum-NLO}.

\subsection{NLO impact factor from dressed self-energies and vertex corrections}

Finally, the virtual corrections with the gluon crossing the shock-wave labeled $\rm SE1$, $\overline{\rm SE1}$\footnote{Excluding their UV singular pieces, which have already been included in Eq.\,\eqref{eq:combining-real-virtual-3}.}, $\rm V1$ and $\overline{\rm V1}$ contribute as (see Sec.\,3.3 and 3.4 in \cite{Caucal:2023nci})
\begin{align}
    &\left.\frac{\der\sigma^{\gamma_{ \lambda=\rmL}^\star+A\to h_1h_2+X}}{\der^2\ktoneh\der^2\kttwoh\der \eta_{h1}\der\eta_{h2}}\right|_{\rm cross}
    =\alpha_s\alpha_{\rm em}e_f^2\int_0^1\frac{\der \zeta_1}{\zeta_1^2}\int_0^1\frac{\der \zeta_2}{\zeta_2^2}D_{h_2/\bar q}(\zeta_2,\mu_F^2)D_{h_1/q}(\zeta_1,\mu_F^2) \ \nonumber\\
    &\times \ \delta\left(1-z_1-z_2\right)\int\frac{\der^2\bt\der^2\bt'}{(2\pi)^4}  e^{-i\Kt\cdot\rbbpt}G^{ij}_{z_f}(\rbbpt)\nonumber\\
    &\times \left\{\frac{2\alpha_sC_F}{\pi}\Hcal_{\rm NLO,3}^{\lambda=\rmL,ij}(\Pt,Q,z_1,z_2)-\frac{2\alpha_s}{\pi}\Hcal_{\rm NLO,4}^{\lambda=\rmL,ij}(\Pt,Q,z_1,z_2)\right\}+(h_1\leftrightarrow h_2) \,, \label{eq:NLO-cross}
\end{align}
where the hard coefficient function for the dressed self-energy is given by
\begin{align}
    \Hcal_{\rm NLO,3}^{\lambda=\rmL,ij}(\Pt)&=\Hcal^{\lambda=\rmL,ij}_{\rm LO}(\Pt)\left[\frac{1}{2}-\frac{\pi^2}{6}-\frac{3}{4}\ln\left(\frac{z_2(1+\chi^2)}{\chi^2}\right)\right]\nonumber\\
    &+\Hcal^{\lambda=\rmL,ij}_{\rm LO}(\Pt)C_F\int_{z_f}^{z_1}\frac{\der z_g}{z_g}\left[-1+\ln\left(\frac{1+\chi^2}{\chi^2}\right)-\ln\left(\frac{z_g}{z_2z_1}\right)\right]\,, \label{eq:HNLO3-final}
\end{align}
and the hard coefficient function for the dressed vertex correction
\begin{align}
      &\Hcal^{\lambda=\rmL,ij}_{\rm NLO,4}(\Pt)=\Hcal_{\rm LO}^{ij,\lambda=\textrm{L}}(\Pt)\times\left\{\frac{N_c}{2}\left[-\frac{1}{8}+\frac{1}{8(z_2-z_1\chi^2)}-\frac{1}{2}\textrm{Li}_2\left(\frac{z_2-z_1\chi^2}{z_2(1+\chi^2)}\right)\right.\right.\nonumber\\
       &\left.+\left(\frac{3}{8}-\frac{z_2-2\chi^2+z_1\chi^4}{8(z_2-z_1\chi^2)^2}\right)\ln\left(\frac{z_2(1+\chi^2)}{\chi^2}\right)\right]+\frac{1}{2N_c}\left[-\frac{\pi^2}{24}-\frac{3}{8}+\frac{3}{8(z_2-z_1\chi^2)}\right.\nonumber\\
       &\left.+\textrm{Li}_2\left(\frac{z_2-z_1\chi^2}{z_2(1+\chi^2)}\right)\left.+\left(\frac{9}{8}-\frac{3z_2+2z_2\chi^2+z_1\chi^4}{8(z_2-z_1\chi^2)^2}\right)\ln\left(\frac{z_2(1+\chi^2)}{\chi^2}\right)+\frac{1}{4}\textrm{Li}_2\left(\frac{z_1\chi^2-z_2}{\chi^2}\right)\right]\right\}\nonumber\\
       &+\Hcal^{\lambda=\rmL,ij}_{\rm LO}(\Pt)C_F\int_{z_f}^{z_1}\frac{\der z_g}{z_g}\left[-1+\ln\left(\frac{1+\chi^2}{\chi^2}\right)-\ln\left(\frac{z_g}{z_2z_1}\right)\right]\nonumber\\
       &+\Hcal^{\lambda=\rmL,ij}_{\rm LO}(\Pt)\frac{N_c}{2}\left[\ln\left(\frac{z_1}{z_f}\right)-\frac{1}{4}\ln^2\left(\frac{z_1}{z_f}\right)-\frac{1}{2}\ln\left(\frac{z_1}{z_f}\right)\ln\left(\frac{z_2(1+\chi^2)}{\chi^2}\right)\right]\,.\label{eq:HNLO4-final}
\end{align}
For these hard coefficient functions to be well defined, the subtraction term to regulate the rapidity divergence as $z_g\to 0$ has been amended by a kinematic constraint. In turn, the rapidity dependence of the WW gluon TMD must be obtained from a kinematically constrained B-JIMWLK equation \cite{Caucal:2023nci}. 

Although not obvious at first sight, Eq.\,\eqref{eq:NLO-cross} contains a Sudakov double logarithm in the hard factor associated with the dressed vertex correction. Indeed, in the last line of Eq.\,\eqref{eq:HNLO4-final} proportional to $N_c/2$, the $\ln^2(z_f)$ term generates a Sudakov double logarithm since $z_f\sim \Kt^2/\Pt^2$, while the $C_F$ dependent Sudakov in the undone $z_g$ integral cancels between Eq.\,\eqref{eq:HNLO3-final} and Eq.\,\eqref{eq:HNLO4-final}. In the end, the overall coefficient of the double logarithm is
\begin{align}
    \frac{\alpha_s N_c}{2\pi}\ln^2\left(\frac{\Pt^2}{\Kt^2}\right) \,,
\end{align}
as discussed around Eq.\,\eqref{eq:cross-Sud-heuristic} in the previous section.

\subsection{Sum of all contributions to the NLO impact factor}
\label{sub:sum-NLO}

Finally, we combine all the different contributions computed in the previous subsections. First, we consider the terms which are $\mu_F$ and $z_f$ dependent, as they will build the resulting factorization scale dependence of the NLO impact factor $\mathcal{I}(\Pt,Q^2,z_1,z_2;z_f,\mu_F)$. To do so, we rewrite the rapidity factorization scale $z_f$ defined in terms of $k_f^-$ as a factorization scale in longitudinal momentum fraction $x_f$ with respect to the large $P^+$ component of the target's four-momentum. This is done consistently both in the NLO impact factor and in the small-$x$ evolution equation for the WW gluon TMD via the change of variable
\begin{align}
    \ln(z_f)=-\ln\left(\frac{\Pt^2\rbbpt^2}{c_0^2}\right)+1-\ln\left(\frac{1+\chi^2}{z_1z_2}\right)+\ln\left(\frac{x_c}{x_f}\right)\,, \label{eq:zf-to-xf}
\end{align}
where the reference $x_c$ scale is conveniently defined as
\begin{align}
    x_c=\frac{x_g}{ec_0^2}\,.
\end{align}
The numerical prefactor $(ec_0^2)^{-1}$ is chosen to simplify the NLO coefficient functions. The WW gluon TMD is thus evolved using the $\eta$-ordered kinematically constrained B-JIWMLK equation~\cite{Ducloue:2019ezk} (see also supplemental material 3 of \cite{Caucal:2023fsf} for the specific case of the WW gluon TMD) from some initial condition at $\eta_0=\ln(1/x_0)$ --- with $x_0$ typically of order 0.01 --- up to $\eta_f=\ln(1/x_f)$.

Since $x_f$ should be chosen of the order of $x_g$ for our particular process, one sees that besides the first term in Eq.\,\eqref{eq:zf-to-xf}, all other terms are $\mathcal{O}(1)$. We have then $z_f\sim 1/(\Pt^2\rbbpt^2)\sim \Kt^2/\Pt^2$ as explained in section~\ref{sec:heuristic-discussion}. In addition, since the rapidity dependence of the WW gluon TMD is determined by the kinematically constrained B-JIMWLK evolution equation, we must include a correction term to the NLO coefficient function from the collinearly singular diagrams computed in subsection~\ref{sub:coll-frag}. Indeed, in the calculation of section~\ref{sub:coll-frag}, we have subtracted the rapidity divergence as $z_g\to 0$ without kinematic constraint. To have a consistent matching between the NLO impact factor and the kinematically constrained rapidity evolution of the WW TMD (see also \cite{Beuf:2014uia} for a similar procedure in the context of the NLO calculation of the fully inclusive DIS cross-section at small-$x$), one must add the following finite piece to the NLO impact factor, corresponding to the integral of the BK kernel with kinematic constraint:
\begin{align}
    I_{\rm kc}(X)&\equiv \int_0^1\frac{\der \xi}{\xi}\int\frac{\der^2 \boldsymbol{\hat r_\perp} }{\pi} \frac{1}{\boldsymbol{\hat r_\perp}^2(\boldsymbol{\hat r_\perp}+\boldsymbol{\hat n_\perp})^2}\Theta\left(\textrm{min}(\boldsymbol{\hat r_\perp}^2,(\boldsymbol{\hat r_\perp}+\boldsymbol{\hat n_\perp})^2)-\frac{1}{\xi X}\right)\\
    &= \left\{
\begin{array}{ll}
 \ln^2(X) +2\textrm{Li}_2\left(\frac{1}{X}\right)&  \quad \quad X \ge 4 \\
 \frac{1}{\pi}\int_{X^{-1/2}}^{\infty}\der u \, \frac{8}{u(u^2-1)}\arctan\left(\frac{u-1}{u+1}\sqrt{\frac{2u+1}{2u-1}}\right)\ln(u^2X) & \quad \quad  0 < X < 4 \,,
\end{array}
\right.
\end{align}
where $\boldsymbol{\hat n_\perp}$ is an arbitrary unit vector in the transverse plane. A detailed derivation of the second equality is given in the appendix (F2) of \cite{Caucal:2022ulg}.

Concerning the transverse momentum factorization scale $\mu_F$, we have $\mu_F\sim K_\perp\sim c_0/r_{bb'}$ so we replace
\begin{align}
    \ln(\mu_F^2)=\ln\left(\frac{c_0^2}{\rbbpt^2}\right)+\ln\left(\frac{\mu_F^2\rbbpt^2}{c_0^2}\right)\,,
\end{align}
wherever $\mu_F^2$ does not appear in combination with $\rbbpt^2$.
The result of this calculation for each (parton-level) Fourier mode (recall Eqs.\,\eqref{eq:Fourier-dec}-\eqref{eq:hadron-Fourier-dec}) is
\begin{align}
        &\left.\frac{\der\sigma^{\gamma_{ \lambda=\rmL}^\star+A\to h_1h_2+X,(0)}}{\der^2\ktoneh\der^2\kttwoh\der \eta_{h1}\der\eta_{h2}}\right|_{\rm NLO}
    =\alpha_s\alpha_{\rm em}e_f^2\int_0^1\frac{\der \zeta_1}{\zeta_1^2}\int_0^1\frac{\der \zeta_2}{\zeta_2^2}D_{h_2/\bar q}(\zeta_2,\mu_F^2)D_{h_1/q}(\zeta_1,\mu_F^2) \ \nonumber\\
    &\times \ \delta\left(1-z_1-z_2\right)\Hcal_{\rm LO}^{\lambda=\rmL,0}(\Pt,Q,z_1,z_2)\int\frac{\der^2\bt\der^2\bt'}{(2\pi)^4}  e^{-i\Kt\cdot\rbbpt}\hat G^{0}_{\eta_f}(\rbbpt)\nonumber\\
    &\times\frac{\alpha_s}{\pi}\left\{-\left[\frac{C_F}{2}+\frac{N_c}{4}\right]\ln^2\left(\frac{\Pt^2\rbbpt^2}{c_0^2}\right)+\left[C_F\left(\ln(z_1z_2)+\frac{3}{2}\right)-N_c\ln(1+\chi^2)\right]\ln\left(\frac{\Pt^2\rbbpt^2}{c_0^2}\right)\right.\nonumber\\
    &-\frac{1}{2}\left[\frac{\mathcal{K}_{\rm DGLAP}\otimes D_{h_1/q}\left(\zeta_1,\mu_F^2\right)}{D_{h_1/q}\left(\zeta_1,\mu_F^2\right)}+\frac{\mathcal{K}_{\rm DGLAP}\otimes D_{h_2/q}\left(\zeta_2,\mu_F^2\right)}{D_{h_2/q}\left(\zeta_2,\mu_F^2\right)}\right]\ln\left(\frac{\mu_F^2\rbbpt^2}{c_0^2}\right)\nonumber\\
    &\left.+\frac{N_c}{2}\ln^2\left(\frac{x_c}{x_f}\right)+N_c\ln\left(\frac{x_c}{x_f}\right)-\frac{N_c}{2}I_{\rm kc}(x_c/x_f)+\mathcal{C}^{\lambda=\rmL}_{\overline{\rm MS}}(\chi,z_1,z_2)\right\}\nonumber\\
    &+\alpha_s\alpha_{\rm em}e_f^2 \frac{\alpha_s}{2\pi N_c}\left[1-\ln(z_1z_2)\right]\int_0^1\frac{\der \zeta_1}{\zeta_1^2}\int_0^1\frac{\der \zeta_2}{\zeta_2^2}D_{h_2/\bar q}(\zeta_2,\mu_F^2)D_{h_1/q}(\zeta_1,\mu_F^2) \ \nonumber\\
    &\times \ \delta\left(1-z_1-z_2\right)\Hcal_{\rm LO}^{\lambda=\rmL,0}(\Pt,Q,z_1,z_2)\int\frac{\der^2\bt\der^2\bt'}{(2\pi)^4}  e^{-i\Kt\cdot\rbbpt}\hat h^{0}_{z_f}(\rbbpt)
    \,,\label{eq:NLO-impact-factor-final}
\end{align}
for the contribution coming from the $n=0$ parton-level Fourier harmonic, and 
\begin{align}
    &\left.\frac{\der\sigma^{\gamma_{ \lambda=\rmL}^\star+A\to h_1h_2+X,(2)}}{\der^2\ktoneh\der^2\kttwoh\der \eta_{h1}\der\eta_{h2}}\right|_{\rm NLO}
    =\alpha_s\alpha_{\rm em}e_f^2\int_0^1\frac{\der \zeta_1}{\zeta_1^2}\int_0^1\frac{\der \zeta_2}{\zeta_2^2}D_{h_2/\bar q}(\zeta_2,\mu_F^2)D_{h_1/q}(\zeta_1,\mu_F^2) \ \nonumber\\
    &\times \ \cos(2\phi)\delta\left(1-z_1-z_2\right)\Hcal_{\rm LO}^{\lambda=\rmL,0}(\Pt,Q,z_1,z_2)\int\frac{\der^2\bt\der^2\bt'}{(2\pi)^4}  e^{-i\Kt\cdot\rbbpt}\cos(2\Phi)\hat h^{0}_{\eta_f}(\rbbpt)\nonumber\\
    &\times\frac{\alpha_s}{\pi}\left\{-\left[\frac{C_F}{2}+\frac{N_c}{4}\right]\ln^2\left(\frac{\Pt^2\rbbpt^2}{c_0^2}\right)+\left[C_F\left(\ln(z_1z_2)+\frac{3}{2}\right)-N_c\ln(1+\chi^2)\right]\ln\left(\frac{\Pt^2\rbbpt^2}{c_0^2}\right)\right.\nonumber\\
    &-\frac{1}{2}\left[\frac{\mathcal{K}_{\rm DGLAP}\otimes D_{h_1/q}\left(\zeta_1,\mu_F^2\right)}{D_{h_1/q}\left(\zeta_1,\mu_F^2\right)}+\frac{\mathcal{K}_{\rm DGLAP}\otimes D_{h_2/q}\left(\zeta_2,\mu_F^2\right)}{D_{h_2/q}\left(\zeta_2,\mu_F^2\right)}\right]\ln\left(\frac{\mu_F^2\rbbpt^2}{c_0^2}\right)\nonumber\\
    &\left.+\frac{N_c}{2}\ln^2\left(\frac{x_c}{x_f}\right)+N_c\ln\left(\frac{x_c}{x_f}\right)-\frac{N_c}{2}I_{\rm kc}(x_c/x_f)+\mathcal{C}^{\lambda=\rmL}_{\overline{\rm MS}}(\chi,z_1,z_2)+\frac{1}{2N_c}\left[-\frac{5}{4}+\frac{1}{2}\ln(z_1z_2)\right]\right\}\nonumber\\
    &+\alpha_s\alpha_{\rm em}e_f^2\int_0^1\frac{\der \zeta_1}{\zeta_1^2}\int_0^1\frac{\der \zeta_2}{\zeta_2^2}D_{h_2/\bar q}(\zeta_2,\mu_F^2)D_{h_1/q}(\zeta_1,\mu_F^2) \cos(2\phi)\delta\left(1-z_1-z_2\right)\ \nonumber\\
    &\times \ \Hcal_{\rm LO}^{\lambda=\rmL,0}(\Pt,Q,z_1,z_2)\int\frac{\der^2\bt\der^2\bt'}{(2\pi)^4}  e^{-i\Kt\cdot\rbbpt}\cos\left(2 \Phi\right)\hat G^{0}_{z_f}(\rbbpt) \frac{\alpha_s}{\pi N_c}\left[1-\ln(z_1z_2)\right]
    \,,\label{eq:NLO-impact-factor-final-n=2}
\end{align}
for the contribution coming from the $n=2$ parton-level Fourier harmonic. The contributions to the hadron-level cross-section coming from higher order Fourier components with $n = 2p$ and $p \ge 2$ read
\begin{align}
        &\left.\frac{\der\sigma^{\gamma_{ \lambda=\rmL}^\star+A\to h_1h_2+X,(n=2p)}}{\der^2\ktoneh\der^2\kttwoh\der \eta_{h1}\der\eta_{h2}}\right|_{\rm NLO}
    =\alpha_s\alpha_{\rm em}e_f^2\int_0^1\frac{\der \zeta_1}{\zeta_1^2}\int_0^1\frac{\der \zeta_2}{\zeta_2^2}D_{h_2/\bar q}(\zeta_2,\mu_F^2)D_{h_1/q}(\zeta_1,\mu_F^2) \ \nonumber\\
    &\times 2\cos(n\phi) \delta\left(1-z_1-z_2\right)\Hcal_{\rm LO}^{\lambda=\rmL,0}(\Pt,Q,z_1,z_2)\int\frac{\der^2\bt\der^2\bt'}{(2\pi)^4}  e^{-i\Kt\cdot\rbbpt}\cos\left(n \Phi\right)\hat G^{0}_{\eta_f}(\rbbpt)\nonumber\\
    &\times\frac{2\alpha_s(-1)^{p+1}}{n\pi N_c}\left\{\sum_{k=1}^p\frac{1}{k}-\frac{1}{n}-\frac{1}{2}\ln(z_1z_2)\right\}\nonumber\\
    &+\alpha_s\alpha_{\rm em}e_f^2 \int_0^1\frac{\der \zeta_1}{\zeta_1^2}\int_0^1\frac{\der \zeta_2}{\zeta_2^2}D_{h_2/\bar q}(\zeta_2,\mu_F^2)D_{h_1/q}(\zeta_1,\mu_F^2)2\cos(n\phi) \ \nonumber\\
    &\times \ \delta\left(1-z_1-z_2\right)\Hcal_{\rm LO}^{\lambda=\rmL,0}(\Pt,Q,z_1,z_2)\int\frac{\der^2\bt\der^2\bt'}{(2\pi)^4}  e^{-i\Kt\cdot\rbbpt}\cos\left(n\Phi\right)\hat h^{0}_{z_f}(\rbbpt)\nonumber\\
    &\times\frac{\alpha_s(-1)^{p}}{(n^2-4)\pi N_c}\left\{(n+2)\sum_{k=1}^{p-1}\frac{1}{k}+(n-2)\sum_{k=1}^{p+1}\frac{1}{k}-\frac{2(n^2+4)}{n^2-4}-n\ln(z_1z_2)\right\}
    \,.\label{eq:NLO-impact-factor-final-n=2p}
\end{align}
The $n\ge 4$ harmonics in the Fourier decomposition are non-zero at NLO because of the real final state interferences between the quark and the antiquark. Nevertheless, they are sub-leading in the large $N_c$ limit.

We now comment on the various terms inside the curly bracket in Eq.\,\eqref{eq:NLO-impact-factor-final} and Eq.\,\eqref{eq:NLO-impact-factor-final-n=2}. The first term is the Sudakov double log, which naturally appears in terms of $\rbbpt$ conjugate to $\Kt$. The value of the coefficient confirms the qualitative calculation in section~\ref{sec:heuristic-discussion}. The second term is a Sudakov single logarithm which also differs from the inclusive dijet case~\cite{Caucal:2023fsf}. The simple result for the coefficient of this single log comes from highly nontrivial cancellations between all the relevant diagrams. In particular, its independence with respect to $x_f$ nor $\mu_F$ is a crucial consistency check of our calculation. One also observes that this single log depends on the "quark anomalous dimension" $\frac{\alpha_sC_F}{\pi}\frac{3}{4}$ which will be discussed in the next section. The Sudakov logarithms displayed in the third line of Eqs.\,\eqref{eq:NLO-impact-factor-final}-\eqref{eq:NLO-impact-factor-final-n=2} are identical for the $n=0$ and $n=2$ parton-level Fourier harmonics which are non-vanishing at LO.

The terms proportional to $\ln(\mu_F^2\rbbpt^2/c_0^2)$ in Eqs.\,\,\eqref{eq:NLO-impact-factor-final} and \eqref{eq:NLO-impact-factor-final-n=2} ensure that the cross-section is $\mu_F$ independent up to powers of $\alpha_s^2$ corrections. The $x_f$ dependent terms are the leftover of the rapidity evolution once formulated with plus light-cone ordering. Note that the $\eta_f=\ln(1/x_f)$ dependence of the WW gluon TMD $\hat G^{ij}_{\eta_f}(\rbbpt)$ cannot be obtained from the $x_f$ dependence of the NLO impact factor. Indeed, the NLO impact factor corresponding to gluons with $z_g\ge z_f$ has undergone a power expansion in $K_\perp/P_\perp$, while the small-$x$ phase space associated with gluons with $z_g\le z_f$ cannot be expanded since the power corrections $(K_\perp/P_\perp)^n$ are accompanied by powers $1/z_g^n$. These power corrections in the small-$x$ phase space are effectively absorbed into the non-linear evolution equation of the WW gluon TMD ~\cite{Caucal:2023nci}. Instead, the $\eta_f$ dependence is given by a kinematically constrained B-JIMWLK evolution applied to the WW correlator along the lines of appendix~(3) in \cite{Caucal:2023fsf}.

Finally, $\mathcal{C}^\lambda_{\overline{\rm MS}}$ is the NLO coefficient function in the $\overline{\rm MS}$-scheme which does not contain any Sudakov logarithm. For a longitudinally polarized virtual photon, it reads
\begin{align}
    \mathcal{C}_{\overline{\rm MS}}^{\lambda=\rmL}(\chi,z_1,z_2)&=\frac{N_c}{2}\left[3-\frac{5\pi^2}{6}-\frac{3}{2}\ln\left(\frac{z_1z_2}{\chi^2}\right)-\ln\left(\frac{1+\chi^2}{z_1}\right)\ln\left(\frac{1+\chi^2}{z_2}\right)\right.\nonumber\\
    &+\left\{-\frac{1}{4(z_2-z_1\chi^2)}+\frac{(1+\chi^2)(z_2(2z_2-z_1)+z_1(2z_1-z_2)\chi^2)}{4(z_2-z_1\chi^2)^2}\ln\left(\frac{z_2(1+\chi^2}{\chi^2}\right)\right.\nonumber\\
    &\left.\left.+\textrm{Li}_2\left(\frac{z_2-z_1\chi^2}{z_2(1+\chi^2)}\right)+(1\leftrightarrow2)\right\}\right]\nonumber\\
    &+\frac{1}{2N_c}\left[-2+\frac{11\pi^2}{12}-\ln(\chi)+\frac{3}{2}\ln(z_1z_2)-\frac{3}{4}\ln^2\left(\frac{z_1}{z_2}\right)\right.\nonumber\\
    &+\left\{\frac{1}{4(z_2-z_1\chi^2)}+\frac{(1+\chi^2)z_1(z_2-(1+z_1)\chi^2)}{4(z_2-z_1\chi^2)^2}\ln\left(\frac{z_2(1+\chi^2)}{\chi^2}\right)\right.\nonumber\\
    &\left.\left.+\frac{1}{2}\textrm{Li}_2(z_2-z_1\chi^2)-\frac{1}{2}\textrm{Li}_2\left(\frac{z_2-z_1\chi^2}{z_2}\right)+(1\leftrightarrow2)\right\}\right]\label{eq:C_MS_L} \,.
\end{align}
Note that the coefficient function depends on the Fourier mode. In particular, for the $n=2$ case, one has the substitution $\mathcal{C}_{\overline{\rm MS}}^{\lambda=\rmL}\to \mathcal{C}_{\overline{\rm MS}}^{\lambda=\rmL}+\frac{1}{2N_c}\left[-\frac{5}{4}+\frac{1}{2}\ln(z_1z_2)\right]$ due to the real final state interference contribution.

The NLO coefficient function $ \mathcal{C}_{\overline{\rm MS}}^{\lambda=\rmT}$ for a transversely polarized virtual photon is given in appendix~\ref{app:transverse}. These NLO corrections are very similar to those obtained in the dijet case~\cite{Caucal:2023nci,Caucal:2023fsf}, they only differ by a finite term which only depends on the jet radius $R$:
\begin{align}
    \mathcal{C}_{\overline{\rm MS}}^{\lambda}- \mathcal{C}_{\rm jet}^{\lambda}&=C_F\left[-6+\frac{2\pi^2}{3}+\frac{3}{2}\ln(R^2)\right]\,,
\end{align}
an identity valid for both virtual photon polarizations.
By construction, the NLO coefficient function is free of any small-$x$ logarithm (of the form $\alpha_s\ln(1/x_f)$), collinear logarithm (of the form $\ln(\mu_F^2/\Lambda_{\rm QCD}^2)$), nor Sudakov logarithms (of the form $\ln(\Pt^2\rbbpt^2)$). Yet, if additional constraints are imposed on the final state, they can nevertheless develop arbitrarily large logarithmic corrections. For instance, in the asymmetric (or aligned jet) configuration where $z_1\sim 1$ and $z_2\ll 1$ (or vice-versa) with fixed $\chi$. In this regime, one easily sees from the analytic expression Eq.\,\eqref{eq:C_MS_L} that the $\overline{\rm MS}$ coefficient function behaves like
\begin{align}
     \mathcal{C}_{\overline{\rm MS}}^{\lambda}(\chi,z_1,z_2)\underset{z_2\ll 1}{=}-\frac{N_c-C_F}{2}\ln^2(z_2)+\mathcal{O}(\ln(z_2))\,,
\end{align}
for both transversely and longitudinally polarized virtual photons. The new large double logarithm $\ln^2(z_2)$ requires specific resummation, beyond the scope of the present paper.

\section{Sudakov resummation for the TMD fragmentation functions and\\ Weizs\"{a}cker-Williams gluon distribution}
\label{sec:resum}

We argue in this section that the resummation formula for inclusive back-to-back dihadron production in DIS must involve TMD fragmentation functions instead of collinear fragmentation functions.
To consistently resum to all orders the large Sudakov logarithms, whose $\mathcal{O}(\alpha_s)$ term of the full series in $\alpha_s$ has been isolated in the previous section, we introduce transverse-momentum dependent fragmentation functions $\overline{D}_{h_1/q}$ in coordinate space representation and make the replacement
\begin{align}
    D_{h_1/q}(\zeta_1,\mu_F)\to \overline{D}_{h_1/q}(\zeta_1,\rbbpt ; \mu_F, P_\perp) \,,
\end{align}
in our NLO expression obtained in Eq.\,\eqref{eq:NLO-impact-factor-final}.
According to the CSS formalism, the $P_\perp$ dependence of the quark TMD fragmentation function $\overline{D}_{h_1/q}$ in the perturbative regime and the fixed QCD coupling approximation reads~\cite{Collins:2011zzd}:
\begin{align}
    \overline{D}_{h_1/q}(\zeta_1,\rbbpt;\mu_F,P_\perp)&\equiv D_{h/q}(\zeta_1,\mu_F)\exp\left(-\frac{\alpha_sC_F}{2\pi}\left[\frac{1}{2}\ln^2\left(\frac{\Pt^2\rbbpt^2}{c_0^2}\right)-\frac{3}{2}\ln\left(\frac{\Pt^2\rbbpt^2}{c_0^2}\right)\right]\right)\,, \label{eq:CSS-TMD}
\end{align}
for $\mu_F= c_0/r_{bb'}$ and likewise for the antiquark into hadron TMD fragmentation function. This expression includes a single Sudakov logarithm related to the finite part $B_q$ of the quark splitting function
\begin{align}
    B_q\equiv\frac{\alpha_sC_F}{2\pi}\frac{3}{2} \,,
\end{align}
which is also present in our NLO impact factor, see Eq.\,\eqref{eq:NLO-impact-factor-final}. The fact that our NLO calculation reproduces both the Sudakov double logarithm and single logarithm proportional to $B_q$ present in the first step of the CSS evolution of the TMD quark fragmentation function is an important cross-check that our interpretation of these large NLO corrections is correct. The strategy to resum all double and single logarithms within a factorized expression is clear: thanks to the introduction of TMD fragmentation functions, the additional double log proportional to $-\alpha_sC_F/(4\pi)$ in Eq.\,\eqref{eq:NLO-impact-factor-final} and the single log proportional to $B_q$ are resummed through the evolution of the TMD fragmentation function and not through the evolution of the WW gluon TMD. The remaining double and single Sudakov logarithms are resummed in a universal soft factor associated with the WW gluon TMD which is identical to the soft factor for inclusive back-to-back dijet production, provided the $R$ dependent single Sudakov logarithms are resummed within jet functions~\cite{Dasgupta:2014yra,Kang:2016mcy,Dai:2016hzf}. In the fixed coupling approximation, the Sudakov factor for the WW TMD in coordinate space reads
\begin{align}
    \Scal(P_\perp,\mu_b)=\exp\left(-\frac{\alpha_s N_c}{\pi}\left[\frac{1}{4}\ln^2\left(\frac{P_\perp^2}{\mu_b^2}\right)+\left(\ln(1+\chi^2)-\frac{C_F}{N_c}\ln(z_1z_2)-\beta_0\right)\ln\left(\frac{P_\perp^2}{\mu_b^2}\right)\right]\right) \,, \label{eq:Sudakov-WW}
\end{align}
with $\mu_b=c_0/r_{bb'}$. The coefficient of the double logarithm is the universal $-\alpha_sN_c/(4\pi)$ factor for gluon TMDs. It is straightforward to generalize Eqs.\,\eqref{eq:CSS-TMD} and \eqref{eq:Sudakov-WW} to the running coupling case. To reach next-to-leading logarithmic accuracy in the resummation, a two-loop running coupling must be employed, for which the Sudakov form factors admit analytic expressions (see e.g. appendix C in \cite{Caucal:2023nci}).

In the end, we can cast our final factorized and resummed NLO cross-section into the following form
\begin{align}
        &\left.\der\sigma^{\gamma_{ \lambda=\rmL}^\star+A\to h_1h_2+X,(0)}\right|_{\rm NLO}
    =\int_0^1\frac{\der \zeta_1}{\zeta_1^2}\int_0^1\frac{\der \zeta_2}{\zeta_2^2}\delta\left(1-z_1-z_2\right)\alpha_s\alpha_{\rm em}e_f^2\Hcal_{\rm LO}^{\lambda=\rmL,0}(\Pt,Q,z_1,z_2)\ \nonumber\\
    &\times \ \int\frac{\der^2\bt\der^2\bt'}{(2\pi)^4}  e^{-i\Kt\cdot\rbbpt}\hat G^{0}_{\eta_f}(\rbbpt)\Scal(P_\perp,\mu_b)\overline{D}_{h_2/\bar q}(\zeta_2,\rbbpt;\mu_F,P_\perp)\overline{D}_{h_1/q}(\zeta_1,\rbbpt;\mu_F,P_\perp) \nonumber\\
    &\times\left\{1-\frac{\alpha_s}{2\pi}\left[\frac{\mathcal{K}_{\rm DGLAP}\otimes D_{h_1/q}\left(\zeta_1,\mu_F^2\right)}{\bar D_{h_1/q}\left(\zeta_1,\mu_F^2\right)}+\frac{\mathcal{K}_{\rm DGLAP}\otimes D_{h_2/q}\left(\zeta_2,\mu_F^2\right)}{\bar D_{h_2/q}\left(\zeta_2,\mu_F^2\right)}\right]\ln\left(\frac{\mu_F^2\rbbpt^2}{c_0^2}\right)\right.\nonumber\\
    &\left.+\frac{\alpha_sN_c}{\pi}\beta_0\ln\left(\frac{\mu_R^2}{P_\perp^2}\right)+\frac{\alpha_sN_c}{2\pi}\ln^2\left(\frac{x_c}{x_f}\right)+\frac{\alpha_sN_c}{\pi}\ln\left(\frac{x_c}{x_f}\right)-\frac{\alpha_sN_c}{2\pi}I_{\rm kc}(x_c/x_f)+\frac{\alpha_s}{\pi}\mathcal{C}^{\lambda=\rmL}_{\overline{\rm MS}}\right\}\nonumber\\
    &+\int_0^1\frac{\der \zeta_1}{\zeta_1^2}\int_0^1\frac{\der \zeta_2}{\zeta_2^2}\delta\left(1-z_1-z_2\right)\alpha_s\alpha_{\rm em}e_f^2\Hcal_{\rm LO}^{\lambda=\rmL,0}(\Pt,Q,z_1,z_2)\frac{\alpha_s}{2\pi N_c}\left[1-\ln(z_1z_2)\right]\ \nonumber\\
    &\times \ \int\frac{\der^2\bt\der^2\bt'}{(2\pi)^4}  e^{-i\Kt\cdot\rbbpt}\hat h^{0}_{\eta_f}(\rbbpt)\Scal(P_\perp,\mu_b)\overline{D}_{h_2/\bar q}(\zeta_2,\rbbpt;\mu_F,P_\perp)\overline{D}_{h_1/q}(\zeta_1,\rbbpt;\mu_F,P_\perp) \,,\label{eq:NLO-impact-factor-final2}
\end{align}
for the $n=0$ parton-level Fourier harmonic, and likewise for the higher order modes given by Eqs.\,\eqref{eq:NLO-impact-factor-final-n=2} and \eqref{eq:NLO-impact-factor-final-n=2p}.
The QCD coupling constant $\alpha_s=\alpha_s(\mu_R)$ is evaluated at the renormalization scale $\mu_R\sim P_\perp$, in particular in the overall $\alpha_s$ prefactor in the first line. 
In this expression, the TMD fragmentation functions appear inside the $\bt$ and $\bt'$ integral as a consequence of the convolution between the final imbalance of the dihadron pair, the momentum transferred from the target, the momentum carried away by soft gluon radiations and the momentum transferred during the fragmentation process. We emphasize that while, strictly speaking, we do not prove here that CSS factorization with both initial and final state TMDs holds to all orders in pQCD at small $x$, it is the first time one demonstrates that the first step in this resummation agrees with an NLO calculation at small $x$. We conjecture that the factorized form of Eq.\,\eqref{eq:NLO-impact-factor-final} is correct to all orders, albeit a two-loop calculation or a more formal proof would be required to be unequivocally convinced. 

\section{Summary and outlook}
\label{sec:outlook}

In this paper, we have computed the NLO corrections to inclusive dihadron production in the Regge limit of DIS within the Color Glass Condensate effective theory. We computed the differential cross-section in the dipole frame and focused on the back-to-back limit (in the transverse plane), the most interesting kinematic regime for gluon saturation searches. In this back-to-back limit, the NLO impact factor is defined as the finite $\mathcal{O}(\alpha_s)$ piece which is left over after isolating the rapidity and collinear divergences in the NLO cross-section. The NLO impact factor contains large double and single logarithmic corrections of the form $\ln(\Pt^2\rbbpt^2)$ with $\rbbpt$ conjugated to the $q\bar q$ transverse momentum imbalance $\Kt$ via Fourier transform. Remarkably, we find that the coefficient of the Sudakov double logarithm for the hadronic final state is given by
\begin{align}
    -\frac{\alpha_s}{2\pi}\left[C_F+\frac{N_c}{2}\right]\ln^2\left(\frac{\Pt^2\rbbpt^2}{c_0^2}\right) \,,
\end{align}
and thus differs from the Sudakov double logarithm for a dijet final state whose coefficient is simply $-\alpha_s N_c/(4\pi)$. Although the coefficient of the Sudakov double logarithm \textit{alone} for inclusive dijet production is smaller than for inclusive dihadron production, the sum of all Sudakov logarithms for dijet and dihadrons are parametrically comparable if the jet radius $R$ is such that $R^2\sim K_\perp/P_\perp$, which for EIC kinematics might be the case, e.g. $K_\perp = 1$ GeV and $P_\perp = 5$ GeV, then $K_\perp/P_\perp \sim 0.2$, while $R^2= 0.16$ for $R=0.4$.

Our final result for the differential cross-section is given by Eq.\,\eqref{eq:NLO-impact-factor-final2}. Our main conclusion is that the differential cross-section factorizes in terms of transverse momentum-dependent fragmentation functions, instead of collinear fragmentation functions. This choice of factorization scheme at small-$x$ allows us to resum the additional $C_F$-dependent Sudakov double logarithms into the CSS evolution of the TMD fragmentation function so that the Sudakov soft factor associated with the initial state WW gluon TMD remains identical between a hadron and a jet measurement as it should from its universality. We also compute the NLO coefficient function in the $\overline{\rm MS}$ scheme, which corresponds to the genuine $\mathcal{O}(\alpha_s)$ corrections after removing the large small-$x$, collinear and Sudakov logarithms. Thus, our final expression Eq.\,\eqref{eq:NLO-impact-factor-final2} enables one to reach NLO accuracy for the back-to-back dihadron correlations in the photon-gluon initiated channel small-$x$ regime. To reach full NLO accuracy, we also need to consider the photon-quark initiated channel, in a future publication we will demonstrate that this channel emerges when considering the dihadron contribution coming from the kinematic limit in which the quark (or anti-quark) and gluon are back-to-back \cite{Caucal:2024xxx}.

Naturally, the next step of this project is to provide numerical predictions for the suppression of azimuthal correlations in dihadron production in DIS. Although our final result is considerably simpler than the inclusive dihadron cross-section for general kinematics in the CGC, it remains a challenge to evaluate it numerically. Indeed, one should first solve a kinematically constrained NLO JIMWLK evolution \cite{Balitsky:2013fea,Kovner:2013ona,Kovner:2014lca,Lublinsky:2016meo} for the WW gluon TMD, which is well beyond the present state-of-the-art\footnote{Some progress in implementing the kinematic constraint in the JIMWLK evolution equation in the Langevin formulation can be found in \cite{Hatta:2016ujq,Korcyl:2024xph}.}. One way to overcome this difficulty is to rely on the Gaussian approximation \cite{Blaizot:2004wv,Iancu:2011nj,Dominguez:2011br} to write the WW TMD in terms of the dipole gluon distribution so that a kinematically constrained NLO BK equation \cite{Beuf:2014uia,Iancu:2015vea,Iancu:2015joa,Ducloue:2019ezk} can be used if one further takes the mean field approximation. In addition, the factorized differential cross-section requires TMD fragmentation functions, with integrated fragmentation functions evolved via DGLAP as an initial condition to the CSS evolution. Hence, a complete numerical framework including three different QCD evolution equations --- B-JIMWLK, DGLAP, and CSS --- must be set up to compute dihadron azimuthal correlation at small $x$ and NLO accuracy. 

In the future it would be interesting to perform a quantitative analysis for the resolving power to saturation physics of dihadrons compared to dijets in DIS \cite{Caucal:2023fsf}, as well as recently proposed processes involving energy correlators \cite{Liu:2023aqb,Kang:2023oqj}. Furthermore, in the photoproduction limit, our results for transversely polarized can be readily applied to study inclusive dihadron production in ultra-peripheral collisions (UPCs) at RHIC and the LHC, where current studies have focused only on dijet production \cite{Kotko:2017oxg} albeit not at full NLO level. Lastly, our present calculations can be extended to study hadron production in proton-nucleus collisions such as photon-hadron \cite{Benic:2022ixp} and hadron-hadron correlations \cite{Giacalone:2018fbc,Stasto:2018rci} where we also expect the necessity to employ TMD fragmentation functions to properly resum Sudakov logarithms associated with the final state.

\smallskip

\noindent{\bf Acknowledgments.} We are grateful to Edmond Iancu and Feng Yuan for valuable discussions. We thank the Nuclear Theory Group at the Lawrence Berkeley National Lab for their hospitality during the completion of this project. F.S. is supported by the Institute for Nuclear Theory’s U.S. DOE under Grant No. DE-FG02-00ER41132. This work is also supported by the U.S. Department of Energy, Office of Science, Office of Nuclear Physics, within the framework of the Saturated Glue (SURGE) Topical Theory Collaboration.

\appendix

\section{Calculation of the real collinear divergence in dihadron production}
\label{app:DGLAP-frag}

The goal of this appendix is to go through the mathematical steps between Eq.\,\eqref{eq:dijet-NLO-long-R2R2-final} and Eq.\,\eqref{eq:final-coll-divergent-real} in the main text.  After a convolution of Eq.\,\eqref{eq:dijet-NLO-long-R2R2-final} with collinear fragmentation functions, we get
\begin{align}
    &\left.\frac{\der\sigma^{\gamma_{ \lambda=\rmL}^\star+A\to h_1h_2+X}}{\der^2\ktoneh\der^2\kttwoh\der \eta_{h1}\der\eta_{h2}}\right|_{\rm R2\times R2^*}=\frac{\alpha_{\rm em}e_f^2N_c}{(2\pi)^6}\int_{\xt,\xt',\yt,\yt'}\Xi_{\rm LO}(\xt,\yt;\xt',\yt') \nonumber\\
    &\times\frac{\alpha_sC_F}{\pi^2}\int_0^1\frac{\der \zeta_1}{\zeta_1^2}\int_0^1\frac{\der \zeta_2}{\zeta_2^2}D^0_{h_1/q}(\zeta_1)D^0_{h_2/\bar q}(\zeta_2) \ 8z_1^3z_2^3Q^2e^{-i\ktone\cdot\rxxtp-i\kttwo\cdot\ryytp}\nonumber\\
    &\times\int_0^1\frac{\der z_g}{z_g}\int\der^2\kgt e^{-i\kgt \cdot\rxxtp}\left\{\delta(1-z_1-z_2-z_g)\frac{(1-z_{2})^2}{z_1^2}\left(1+\frac{z_g}{z_1}+\frac{z_g^2}{2z_1^2}\right)\right.\nonumber\\
    &\left.\times \frac{K_0(\bar Q_{\mathrm{R}2}r_{xy})K_0(\bar Q_{\mathrm{R}2}r_{x'y'})}{\left(\kgt-\frac{z_g}{z_1}\ktone\right)^2}-\delta(1-z_1-z_2)\Theta(z_f-z_g)\frac{K_0(\bar QKr_{xy})K_0(\bar Qr_{x'y'})}{\kgt^2}\right\}\,,
\end{align}
where we recall that $z_1, z_2, \ktone$ and $\kttwo$ are implicit functions of $\zeta_1$, $\zeta_2$ and the external hadronic variables since $\zeta_i=k_{h,i}^{\mu}/k_i^\mu$. After the change of variable from $z_g$ to the variable $\xi$ defined by
\begin{align}
    \xi=\frac{z_1}{z_1+z_g}\,, \label{eq:xi}
\end{align}
representing the longitudinal momentum fraction of the final state quark with respect to the parent quark, we get
\begin{align}
    &\left.\frac{\der\sigma^{\gamma_{ \lambda=\rmL}^\star+A\to h_1h_2+X}}{\der^2\ktoneh\der^2\kttwoh\der \eta_{h1}\der\eta_{h2}}\right|_{\rm R2\times R2^*}=\frac{\alpha_{\rm em}e_f^2N_c}{(2\pi)^6}\int_{\xt,\xt',\yt,\yt'}\Xi_{\rm LO}(\xt,\yt;\xt',\yt') \nonumber\\
    &\times\frac{\alpha_sC_F}{\pi^2}\int_0^1\frac{\der \zeta_1}{\zeta_1^2}\int_0^1\frac{\der \zeta_2}{\zeta_2^2}D^0_{h_1/q}(\zeta_1)D^0_{h_2/\bar q}(\zeta_2) \ 8z_1^3z_2^3Q^2e^{-i\ktone\cdot\rxxtp-i\kttwo\cdot\ryytp}\nonumber\\
    &\times\int\frac{\der \xi}{\xi(1-\xi)}\int\der^2\kgt\left\{\delta\left(1-\frac{z_1}{\xi}-z_2\right)\frac{1+\xi^2}{2\xi^4}K_0\left(\sqrt{\frac{z_2z_1}{\xi}}Qr_{xy}\right)K_0\left(\sqrt{\frac{z_2z_1}{\xi}}Qr_{x'y'}\right)\right.\nonumber\\
    &\left.\times \frac{e^{-i\kgt \cdot\rxxtp}}{\left(\kgt-\frac{1-\xi}{\xi}\ktone\right)^2}-\delta(1-z_1-z_2)\Theta\left(\xi-\frac{z_1}{z_1+z_f}\right)K_0(\bar Qr_{xy})K_0(\bar Q r_{x'y'})\frac{e^{-i\kgt \cdot\rxxtp}}{\kgt^2}\right\}\,.\label{eq:factorization-step0}
\end{align}
The boundaries for the $\xi$ integral will be specified in the next equation. For the first term inside the curly bracket, we then make the change of variable $\zeta_1'=\xi \zeta_1$ and $\kgt'=\kgt-(1-\xi)/\xi \ktone$. The latter change of variables brings an additional phase
\begin{align}
    e^{-i\frac{1-\xi}{\xi}\ktone\cdot\rxxtp} \,, \label{eq:nophase!}
\end{align}
which might be worrisome. As a matter of fact, after the change of variable  $\zeta_1'=\xi \zeta_1$, this is exactly the phase needed to get the expected phase $e^{-i\ktoneh\cdot\rxxtp/\zeta_1'}$ since
\begin{equation}
    \frac{\xi\ktoneh}{\zeta_1'}+\frac{1-\xi}{\xi}\frac{\xi\ktoneh}{\zeta_1'}=\frac{\ktoneh}{\zeta_1'} \,.
\end{equation}
Note that in the calculation performed in \cite{Bergabo:2022tcu}, the phase Eq.\,\eqref{eq:nophase!} remains in their final result, in contrast to our result.

The integral over $\kgt'$ in dimensional regularization yields
\begin{align}
    \mu^\varepsilon\int\frac{\der^{2-\varepsilon} \kgt'}{(2\pi)^{2-\varepsilon}}\frac{e^{-i\kgt'\cdot\rxxtp}}{\kgt'^2}=-\frac{1}{4\pi}\left[\frac{2}{\varepsilon}+\ln(\pi\mu^2e^{\gamma_E}\rxxtp^2)+\mathcal{O}(\varepsilon)\right] \,,
\end{align}
so that we get
\begin{align}
    &\left.\frac{\der\sigma^{\gamma_{ \lambda=\rmL}^\star+A\to h_1h_2+X}}{\der^2\ktoneh\der^2\kttwoh\der \eta_{h1}\der\eta_{h2}}\right|_{\rm R2\times R2^*}=\frac{\alpha_{\rm em}e_f^2N_c}{(2\pi)^6}\frac{(-\alpha_s)C_F}{2\pi}\int_{\xt,\xt',\yt,\yt'}\Xi_{\rm LO}(\xt,\yt;\xt',\yt') \nonumber\\
    &\times\int_0^1\frac{\der \zeta_1'}{\zeta_1'^2}\int_0^1\frac{\der \zeta_2}{\zeta_2^2}D^0_{h_2/\bar q}(\zeta_2) \ 8z_1^3z_2^3Q^2K_0(\bar Qr_{xy})K_0(\bar Qr_{x'y'})e^{-i\ktone\cdot\rxxtp-i\kttwo\cdot\ryytp}\delta\left(1-z_1-z_2\right)\nonumber\\
    &\times\left\{\int_{\zeta_1'}^1\frac{\der \xi}{\xi}\frac{1+\xi^2}{(1-\xi)}D^0_{h_1/q}(\zeta_1'/\xi)-\int_{\frac{z_1}{z_1+z_f}}^1\frac{\der\xi}{\xi}\frac{2}{1-\xi}D^0_{h_1/q}(\zeta_1')\right\}\left[\frac{2}{\varepsilon}+\ln(\pi\mu^2e^{\gamma_E}\rxxtp^2)\right] \,,\label{eq:factorization-step1}
\end{align}
where now, we redefine the variables: $z_1=z_{h1}/\zeta_1'$ and $\ktone=\ktoneh/\zeta_1'$.
Lastly, we introduce the (regularized) quark-quark splitting function,
\begin{align}
    P_{qq}(\xi)\equiv C_F\left[ \frac{1+\xi^2}{(1-\xi)_+}+\frac{3}{2}\delta(1-\xi)\right]\,,
\end{align}
where the plus prescription is defined as
\begin{align}
    (f(x))_{+} = f(x) -\delta(1-x) \int_0^1 \der z f(z) \,.
\end{align}
One can then use the identity \cite{Bergabo:2022tcu}
\begin{align}
    \int_{\zeta_1}^1\frac{\der \xi}{\xi}\frac{1+\xi^2}{(1-\xi)}D^0_{h_1/q}(\zeta_1/\xi)=\int_{\zeta_1}^1\frac{\der \xi}{\xi}\frac{P_{qq}(\xi)}{C_F}D^0_{h_1/q}(\zeta_1/\xi)+\int_0^1\der\xi\frac{1+\xi^2}{1-\xi}D^0_{h_1/q}(\zeta_1)\,,
\end{align}
to express the first term inside the curly bracket in Eq.\,\eqref{eq:factorization-step1} in terms of the quark DGLAP splitting function plus a remainder which is combined with the second term depending on $z_f$ inside the curly bracket in Eq.\,\eqref{eq:factorization-step1}. The sum of this remainder and the $z_f$ dependent integral has no divergence as $\xi\to 1$:
\begin{align}
    \int_0^1\der\xi\frac{1+\xi^2}{1-\xi}-\int_{\frac{z_1}{z_1+z_f}}^1\frac{\der\xi}{\xi}\frac{2}{1-\xi}=2\ln\left(\frac{z_1}{z_f}\right)-\frac{3}{2}\,.
\end{align}
Combining these results we find
\begin{align}
    &\left.\frac{\der\sigma^{\gamma_{ \lambda=\rmL}^\star+A\to h_1h_2+X}}{\der^2\ktoneh\der^2\kttwoh\der \eta_{h1}\der\eta_{h2}}\right|_{\rm R2\times R2^*}=\frac{\alpha_{\rm em}e_f^2N_c}{(2\pi)^6}\frac{(-\alpha_s)}{2\pi}\int_{\xt,\xt',\yt,\yt'}\Xi_{\rm LO}(\xt,\yt;\xt',\yt') \nonumber\\
    &\times\int_0^1\frac{\der \zeta_1'}{\zeta_1'^2}\int_0^1\frac{\der \zeta_2}{\zeta_2^2}D^0_{h_2/\bar q}(\zeta_2) \ 8z_1^3z_2^3Q^2K_0(\bar Qr_{xy})K_0(\bar Qr_{x'y'})e^{-i\ktone\cdot\rxxtp-i\kttwo\cdot\ryytp}\delta\left(1-z_1-z_2\right)\nonumber\\
    &\times\left\{\int_{\zeta_1'}^1\frac{\der \xi}{\xi} P_{qq}(\xi)D^0_{h_1/q}\left(\frac{\zeta_1'}{\xi}\right)+ C_F \left[ 2\ln\left(\frac{z_1}{z_f}\right)-\frac{3}{2}\right] D^0_{h_1/q}(\zeta_1') \right\}\left[\frac{2}{\varepsilon}+\ln(\pi\mu^2e^{\gamma_E}\rxxtp^2)\right] \,. \label{eq:factorization-step2}
\end{align}
Lastly, after renaming $\zeta_1'\to \zeta_1$, the above result simplifies into Eq.\,\eqref{eq:final-coll-divergent-real}.

\paragraph{The gluon into hadron fragmentation contribution.} So far, we have tagged the hadrons coming from the quark or antiquark fragmentation. At NLO, there is also a contribution associated with gluon fragmentation. In the case of SIDIS at small $x$, such a term has been studied in the recent paper \cite{Bergabo:2024ivx}. Coming back to Eq.\,\eqref{eq:dijet-NLO-long-R2R2-final} and integrating out the quark instead of the gluon, we find
\begin{align}
    &\left.\frac{\der\sigma^{\gamma_{ \lambda=\rmL}^\star+A\to h_1h_2+X}}{\der^2\ktoneh\der^2\kttwoh\der \eta_{h1}\der\eta_{h2}}\right|_{\rm R2\times R2^*,g}=\frac{\alpha_{\rm em}e_f^2N_c}{(2\pi)^6}\int_{\xt,\xt',\yt,\yt'}\Xi_{\rm LO}(\xt,\yt;\xt',\yt') \nonumber\\
    &\times\frac{\alpha_sC_F}{\pi^2}\int_0^1\frac{\der \zeta_1}{\zeta_1^2}\int_0^1\frac{\der \zeta_2}{\zeta_2^2}D^0_{h_1/g}(\zeta_1)D^0_{h_2/\bar q}(\zeta_2) \ 8z_2^3(1-z_2)^2Q^2e^{-i\ktone\cdot\rxxtp-i\kttwo\cdot\ryytp}\nonumber\\
    &\times\int_0^1\der z_1\delta(1-z_1-z_2-z_g)\left(1+\frac{z_g}{z_1}+\frac{z_g^2}{2z_1^2}\right) K_0(\bar Q_{\mathrm{R}2}r_{xy})K_0(\bar Q_{\mathrm{R}2}r_{x'y'})\nonumber\\
    &\times \int\der^2\ktone \frac{e^{-i\ktone \cdot\rxxtp}}{\left(\kgt-\frac{z_g}{z_1}\ktone\right)^2}\,,
\end{align}
after introducing the bare gluon fragmentation function $D^0_{h_1/g}(\zeta_1)$. Now, the internal partonic variables are related to the measured hadron kinematics according to $z_g=z_{h1}/\zeta_1$ and $\kgt=\ktoneh/\zeta_1$. For this gluon into hadron contribution, one must not subtract any rapidity divergence since the integral over $z_1$ has no singularity as $z_1\to 0$. After the change of variables $z_1\to \xi\equiv z_g/(z_1+z_g)$, we get
\begin{align}
    &\left.\frac{\der\sigma^{\gamma_{ \lambda=\rmL}^\star+A\to h_1h_2+X}}{\der^2\ktoneh\der^2\kttwoh\der \eta_{h1}\der\eta_{h2}}\right|_{\rm R2\times R2^*,g}=\frac{\alpha_{\rm em}e_f^2N_c}{(2\pi)^6}\int_{\xt,\xt',\yt,\yt'}\Xi_{\rm LO}(\xt,\yt;\xt',\yt') \nonumber\\
    &\times\frac{\alpha_sC_F}{\pi^2}\int_0^1\frac{\der \zeta_1}{\zeta_1^2}\int_0^1\frac{\der \zeta_2}{\zeta_2^2}D^0_{h_1/g}(\zeta_1)D^0_{h_2/\bar q}(\zeta_2) \ 8z_2^3z_g^3Q^2e^{-i\kgt\cdot\rxxtp-i\kttwo\cdot\ryytp}\nonumber\\
    &\times\int_{\frac{z_g}{1+z_g}}^1\der \xi \ \delta\left(1-\frac{z_g}{\xi}-z_2\right)\frac{1+(1-\xi)^2}{2\xi^6} K_0\left(\sqrt{\frac{z_2z_g}{\xi}}Qr_{xy}\right)K_0\left(\sqrt{\frac{z_2z_g}{\xi}}Qr_{x'y'}\right)\nonumber\\
    &\times \int\der^2\ktone \frac{e^{-i\ktone \cdot\rxxtp}}{\left(\ktone-\frac{1-\xi}{\xi}\kgt\right)^2} \,.
\end{align}
Finally, integrating over $\ktone$ and rescaling $\zeta_1\to \xi \zeta_1$ yields
\begin{align}
    &\left.\frac{\der\sigma^{\gamma_{ \lambda=\rmL}^\star+A\to h_1h_2+X}}{\der^2\ktoneh\der^2\kttwoh\der \eta_{h1}\der\eta_{h2}}\right|_{\rm R2\times R2^*,g}=\frac{\alpha_{\rm em}e_f^2N_c}{(2\pi)^6}\int_{\xt,\xt',\yt,\yt'}\Xi_{\rm LO}(\xt,\yt;\xt',\yt') \nonumber\\
    &\times\frac{(-\alpha_s)C_F}{2\pi}\int_0^1\frac{\der \zeta_1}{\zeta_1^2}\int_0^1\frac{\der \zeta_2}{\zeta_2^2}D^0_{h_2/\bar q}(\zeta_2) \ 8z_2^3z_g^3Q^2K_0\left(\bar Qr_{xy}\right)K_0\left(\bar Qr_{x'y'}\right)e^{-i\kgt\cdot\rxxtp-i\kttwo\cdot\ryytp}\nonumber\\
    &\times\delta\left(1-z_g-z_2\right)\int_{\zeta_1}^1\frac{\der \xi}{\xi}\frac{1+(1-\xi)^2}{\xi}D^0_{h_1/g}\left(\frac{\zeta_1}{\xi}\right)\left[\frac{2}{\varepsilon}+\ln\left(\pi\mu^2e^{\gamma_E}\rxxtp^2\right)\right] \,.
\end{align}
One recognizes in this expression the $P_{gq}$ splitting function defined by
\begin{align}
    P_{gq}(\xi)\equiv C_F\frac{1+(1-\xi)^2}{\xi}\,,
\end{align}
so that the gluon into hadron contribution factorizes as
\begin{align}
    &\left.\frac{\der\sigma^{\gamma_{ \lambda=\rmL}^\star+A\to h_1h_2+X}}{\der^2\ktoneh\der^2\kttwoh\der \eta_{h1}\der\eta_{h2}}\right|_{\rm R2\times R2^*,g}=\int_0^1\frac{\der \zeta_1}{\zeta_1^2}\int_0^1\frac{\der \zeta_2}{\zeta_2^2}D^0_{h_2/\bar q}(\zeta_2)\left.\frac{\der\sigma^{\gamma_{ \lambda=\rmL}^\star+A\to q\bar q+X}}{\der^2\ktone\der^2\kttwo\der \eta_{1}\der\eta_{2}}\right|_{\rm LO}\nonumber\\
    &\times \frac{(-\alpha_s)}{2\pi}\int_{\zeta_1}^1\frac{\der \xi}{\xi}P_{gq}(\xi)D^0_{h_1/g}\left(\frac{\zeta_1}{\xi}\right)\left[\frac{2}{\varepsilon}+\ln(\pi\mu^2e^{\gamma_E}\rxxtp^2)\right] \,.
\end{align}
This concludes our calculation of the gluon fragmenting into hadron contribution.

\section{NLO coefficient function for transversely polarized virtual photon}
\label{app:transverse}

In this appendix, we report the analytic expression for the NLO coefficient function $\mathcal{C}^{\lambda=\rmT}_{\overline{\rm MS}} $ for back-to-back dihadron production in DIS mediated by a transversely polarized virtual photon. A detailed calculation is provided in the supplemental material of \cite{Caucal:2023fsf}. The inclusive dihadron cross-section in the back-to-back limit is decomposed in Fourier modes according to Eq.\,\eqref{eq:hadron-Fourier-def}. After resummation of the Sudakov logarithms, the $n=0$ component reads
\begin{align}
    &\left.\der\sigma^{\gamma_{ \lambda=\rmT}^\star+A\to h_1h_2+X,(0)}\right|_{\rm NLO}
    =\alpha_s\alpha_{\rm em}e_f^2\int_0^1\frac{\der \zeta_1}{\zeta_1^2}\int_0^1\frac{\der \zeta_2}{\zeta_2^2}\delta\left(1-z_1-z_2\right)\Hcal_{\rm LO}^{\lambda=\rmT,0}\ \nonumber\\
    &\times \ \int\frac{\der^2\bt\der^2\bt'}{(2\pi)^4}  e^{-i\Kt\cdot\rbbpt}\hat G^{0}_{\eta_f}(\rbbpt)\Scal(P_\perp,\mu_b)\overline{D}_{h_2/\bar q}(\zeta_2,\rbbpt;\mu_F,P_\perp)\overline{D}_{h_1/q}(\zeta_1,\rbbpt;\mu_F,P_\perp) \nonumber\\
    &\times\left\{1-\frac{\alpha_s}{2\pi}\left[\frac{\mathcal{K}_{\rm DGLAP}\otimes D_{h_1/q}\left(\zeta_1,\mu_F^2\right)}{\bar D_{h_1/q}\left(\zeta_1,\mu_F^2\right)}+\frac{\mathcal{K}_{\rm DGLAP}\otimes D_{h_2/q}\left(\zeta_2,\mu_F^2\right)}{\bar D_{h_2/q}\left(\zeta_2,\mu_F^2\right)}\right]\ln\left(\frac{\mu_F^2\rbbpt^2}{c_0^2}\right)\right.\nonumber\\
    &\left.+\frac{\alpha_sN_c}{\pi}\beta_0\ln\left(\frac{\mu_R^2}{P_\perp^2}\right)+\frac{\alpha_sN_c}{2\pi}\ln^2\left(\frac{x_c}{x_f}\right)+\frac{\alpha_sN_c}{\pi}\ln\left(\frac{x_c}{x_f}\right)-\frac{\alpha_sN_c}{2\pi}I_{\rm kc}(x_c/x_f)+\frac{\alpha_s}{\pi}\mathcal{C}^{\lambda=\rmT}_{\overline{\rm MS}}\right\}\nonumber\\
    &+\alpha_s\alpha_{\rm em}e_f^2\int_0^1\frac{\der \zeta_1}{\zeta_1^2}\int_0^1\frac{\der \zeta_2}{\zeta_2^2}\delta\left(1-z_1-z_2\right)\Hcal_{\rm LO}^{\lambda=\rmT,0}\left(\frac{-2\chi^2}{1+\chi^4}\right)\frac{\alpha_s}{2\pi N_c}\left[1-\ln(z_1z_2)\right]\ \nonumber\\
    &\times \ \int\frac{\der^2\bt\der^2\bt'}{(2\pi)^4}  e^{-i\Kt\cdot\rbbpt}\hat h^{0}_{\eta_f}(\rbbpt)\Scal(P_\perp,\mu_b)\overline{D}_{h_2/\bar q}(\zeta_2,\rbbpt;\mu_F,P_\perp)\overline{D}_{h_1/q}(\zeta_1,\rbbpt;\mu_F,P_\perp) \,,\label{eq:NLO-impact-factor-final2-T}
\end{align}
with the trace of the LO hard factor defined as usual,
\begin{align}
    \Hcal_{\rm LO}^{\lambda=\rmT,0}(\Pt,Q,z_1,z_2)=\frac{1}{2}\delta^{ij} \Hcal_{\rm LO}^{\lambda=\rmT,ij}(\Pt,Q,z_1,z_2)\,.
\end{align}
For compactness, we write the NLO coefficient function for the azimuthally averaged cross-section ($n=0$ Fourier component in Eq.\,\eqref{eq:Fourier-dec}) as
\begin{align}
    \mathcal{C}^{\lambda=\rmT}_{\overline{\rm MS}}(\chi,z_1,z_2)&=\frac{N_c}{2}f_1^{\lambda=\rmT}(\chi,z_1,z_2)+\frac{1}{2N_c}f_2^{\lambda=\rmT}(\chi,z_1,z_2)\,,\\
    f_1^{\lambda=\rmT}(\chi,z_1,z_2)&=\frac{(1+\chi^2)^2}{1+\chi^4}f_{A,1}^{\lambda=\rmT}-\frac{2\chi^2}{1+\chi^4}f_{B,1}^{\lambda=\rmT} \,, \\
    f_2^{\lambda=\rmT}(\chi,z_1,z_2)&=\frac{(1+\chi^2)^2}{1+\chi^4}f_{A,2}^{\lambda=\rmT}-\frac{2\chi^2}{1+\chi^4}f_{B,2}^{\lambda=\rmT} \,,
\end{align}
where we recall the definition of the kinematic variable $\chi\equiv Q/M_{q\bar q}$. The functions $f_{A,1}$, $f_{A,2}$ are given by
\begin{align}
    &f_{A,1}^{\lambda=\rmT}(\chi,z_1,z_2)=\frac{7}{2}-\frac{5\pi^2}{6}-\frac{3}{2}\ln\left(\frac{z_1z_2}{\chi^2}\right)+\frac{z_1z_2}{z_1^2+z_2^2}-\ln\left(\frac{1+\chi^2}{z_1}\right)\ln\left(\frac{1+\chi^2}{z_2}\right)\nonumber\\
    &+\left\{\frac{z_1^2\chi^2}{2(z_2-z_1\chi^2)(z_1^2+z_2^2)}+\textrm{Li}_2\left(\frac{z_2-z_1\chi^2}{z_2(1+\chi^2)}\right)\right.\nonumber\\
    &-\left.\frac{z_1z_2[4\chi^2-3z_1z_2(1+\chi^2)^2]}{2(z_1^2+z_2^2)(z_2-z_1\chi^2)^2}\ln\left(\frac{z_2(1+\chi^2)}{\chi^2}\right)+(1\leftrightarrow 2)\right\}\,, \label{fA1_def}\\
    &f_{A,2}^{\lambda=\rmT}(\chi,z_1,z_2)=-1+\frac{5\pi^2}{6}+\frac{3}{2}\ln(z_1z_2)-\frac{1}{2}\left[1+\frac{2z_2+z_1^2(1+\chi^2)}{(1+\chi^2)(z_1^2+z_2^2)}\right]\ln^2\left(\frac{z_1}{z_2}\right)\nonumber\\
    &+\frac{1+z_1^2+z_2^2\chi^2}{(1+\chi^2)(z_1^2+z_2^2)}\frac{\pi^2}{6}+\frac{(1-\chi^2)(z_1-z_2)}{2(1+\chi^2)(z_1^2+z_2^2)}\left[\ln\left(\frac{z_2}{z_1}\right)\ln(z_1z_2)+2\textrm{Li}_2\left(-\frac{z_1}{z_2}\right)\right]\nonumber\\
    &-\frac{z_1z_2(1-\chi^2)}{2(z_1-z_2\chi^2)(z_2-z_1\chi^2)(z_1^2+z_2^2)}-\frac{1}{2(1+\chi^2)}\left[1+3\chi^2+\frac{2}{z_1^2+z_2^2}\right]\ln(\chi^2)\nonumber\\
    &+\left\{\frac{2z_2+z_1^2(1+\chi^2)}{(1+\chi^2)(z_1^2+z_2^2)}\left[\textrm{Li}_2(z_2-z_1\chi^2)-\textrm{Li}_2\left(\frac{z_2-z_1\chi^2}{z_2}\right)\right]\right.\nonumber\\
    &\left.+\frac{z_1z_2[-z_2(2+z_1)+2(1+z_1^2)\chi^2-z_1z_2\chi^4]}{2(z_2-z_1\chi^2)^2(z_1^2+z_2^2)}\ln\left(\frac{z_2(1+\chi^2)}{\chi^2}\right)+(1\leftrightarrow 2)\right\} \,, 
\end{align}
and the functions $f_{B,1}$, $f_{B,2}$ are given by
\begin{align}
    &f_{B,1}^{\lambda=\rmT}(\chi,z_1,z_2)=3-\frac{5\pi^2}{6}-\frac{3}{2}\ln\left(\frac{z_1z_2}{\chi^2}\right)-\ln\left(\frac{1+\chi^2}{z_1}\right)\ln\left(\frac{1+\chi^2}{z_2}\right)\nonumber\\
    &+\frac{(1-\chi^2)(z_1z_2-(z_1-z_2)^2\chi^2+z_1z_2\chi^4)}{4\chi^2(z_1-z_2\chi^2)(z_2-z_1\chi^2)(z_1^2+z_2^2)}+\left\{\textrm{Li}_2\left(\frac{z_2-z_1\chi^2}{z_2(1+\chi^2)}\right)\right.\nonumber\\
    &\left.-\frac{z_1z_2(1+\chi^2)(z_2(2z_2-z_1)+z_1(2z_1-z_2)\chi^2)}{2(z_2-z_1\chi^2)^2(z_1^2+z_2^2)}\ln\left(\frac{z_2(1+\chi^2)}{\chi^2}\right)+(1\leftrightarrow2)\right\}\,, \label{fB1_def}\\
    &f_{B,2}^{\lambda=\rmT}(\chi,z_1,z_2)=-\frac{9}{2}+\frac{5\pi^2}{6}+\frac{3}{2}\ln(z_1z_2)+\frac{(1+\chi^2)[z_1z_2(z_2-z_1)^2-(1-2z_1z_2)^2\chi^2+z_1z_2\chi^4]}{4\chi^2(z_1-z_2\chi^2)(z_2-z_1\chi^2)(z_1^2+z_2^2)}\nonumber\\
    &-\left[1+\frac{1}{2(z_1^2+z_2^2)}\right]\ln(\chi^2)+\frac{1-z_1z_2}{z_1^2+z_2^2}\frac{\pi^2}{6}+\frac{-2+3z_1z_2}{2(z_1^2+z_2^2)}\ln^2\left(\frac{z_1}{z_2}\right)\nonumber\\
    &+\left\{\frac{1-z_1z_2}{z_1^2+z_2^2}\left[\textrm{Li}_2\left(z_2-z_1\chi^2\right)-\textrm{Li}_2\left(\frac{z_2-z_1\chi^2}{z_2}\right)\right]\right.\nonumber\\
    &\left.-\frac{z_1^2z_2(1+\chi^2)(z_2-(1+z_1)\chi^2)}{2(z_2-z_1\chi^2)^2(z_1^2+z_2^2)}\ln\left(\frac{z_2(1+\chi^2)}{\chi^2}\right)+(1\leftrightarrow2)\right\} \,.
\end{align}
The $n=2$ Fourier component reads, after Sudakov resummation,
\begin{align}
    &\left.\der\sigma^{\gamma_{ \lambda=\rmT}^\star+A\to h_1h_2+X,(2)}\right|_{\rm NLO}
    =\alpha_s\alpha_{\rm em}e_f^2\int_0^1\frac{\der \zeta_1}{\zeta_1^2}\int_0^1\frac{\der \zeta_2}{\zeta_2^2}\cos(2\phi)\delta\left(1-z_1-z_2\right)\Hcal_{\rm LO}^{\lambda=\rmT,0}\left(\frac{-2\chi^2}{1+\chi^4}\right)\ \nonumber\\
    &\times \ \int\frac{\der^2\bt\der^2\bt'}{(2\pi)^4}  e^{-i\Kt\cdot\rbbpt}\cos(2\Phi)\hat h^{0}_{\eta_f}(\rbbpt)\Scal(P_\perp,\mu_b)\overline{D}_{h_2/\bar q}(\zeta_2,\rbbpt;\mu_F,P_\perp)\overline{D}_{h_1/q}(\zeta_1,\rbbpt;\mu_F,P_\perp) \nonumber\\
    &\times\left\{1-\frac{\alpha_s}{2\pi}\left[\frac{\mathcal{K}_{\rm DGLAP}\otimes D_{h_1/q}\left(\zeta_1,\mu_F^2\right)}{\bar D_{h_1/q}\left(\zeta_1,\mu_F^2\right)}+\frac{\mathcal{K}_{\rm DGLAP}\otimes D_{h_2/q}\left(\zeta_2,\mu_F^2\right)}{\bar D_{h_2/q}\left(\zeta_2,\mu_F^2\right)}\right]\ln\left(\frac{\mu_F^2\rbbpt^2}{c_0^2}\right)\right.\nonumber\\
    &\left.+\frac{\alpha_sN_c}{\pi}\beta_0\ln\left(\frac{\mu_R^2}{P_\perp^2}\right)+\frac{\alpha_sN_c}{2\pi}\ln^2\left(\frac{x_c}{x_f}\right)+\frac{\alpha_sN_c}{\pi}\ln\left(\frac{x_c}{x_f}\right)-\frac{\alpha_sN_c}{2\pi}I_{\rm kc}(x_c/x_f)\right.\nonumber\\
    &\left.+\frac{\alpha_s}{\pi}f_{B}^{\lambda=\rmT}(\chi,z_1,z_2)+\frac{1}{2N_c}\left[-\frac{5}{4}+\frac{1}{2}\ln(z_1z_2)\right]\right\}\nonumber\\
    &+\alpha_s\alpha_{\rm em}e_f^2\int_0^1\frac{\der \zeta_1}{\zeta_1^2}\int_0^1\frac{\der \zeta_2}{\zeta_2^2}\cos(2\phi)\delta\left(1-z_1-z_2\right)\Hcal_{\rm LO}^{\lambda=\rmT,0}\frac{\alpha_s}{\pi N_c}\left[1-\ln(z_1z_2)\right]\ \nonumber\\
    &\times \ \int\frac{\der^2\bt\der^2\bt'}{(2\pi)^4}  e^{-i\Kt\cdot\rbbpt}\cos(2\Phi)\hat G^{0}_{\eta_f}(\rbbpt)\Scal(P_\perp,\mu_b)\overline{D}_{h_2/\bar q}(\zeta_2,\rbbpt;\mu_F,P_\perp)\overline{D}_{h_1/q}(\zeta_1,\rbbpt;\mu_F,P_\perp) \,.\label{eq:NLO-impact-factor-final2-T}
\end{align}
Like for the longitudinally polarized case, the NLO coefficient function for the second harmonic ($n=2$) is given by the substitution 
\begin{align}
   f_{B,2}^{\lambda=\rmT}\to f_{B,2}^{\lambda=\rmT}-\frac{5}{4}+\frac{1}{2}\ln(z_1z_2)\,.
\end{align}
Finally, the $n=2p$, $p\ge3$ Fourier components are given by the following expression,
\begin{align}
        &\left.\der\sigma^{\gamma_{ \lambda=\rmT}^\star+A\to h_1h_2+X,(n=2p)}\right|_{\rm NLO}
    =\alpha_s\alpha_{\rm em}e_f^2\int_0^1\frac{\der \zeta_1}{\zeta_1^2}\int_0^1\frac{\der \zeta_2}{\zeta_2^2}\overline{D}_{h_2/\bar q}(\zeta_2,\rbbpt;\mu_F,P_\perp)\overline{D}_{h_1/q}(\zeta_1,\rbbpt;\mu_F,P_\perp) \ \nonumber\\
    &\times 2\cos(n\phi) \delta\left(1-z_1-z_2\right)\Hcal_{\rm LO}^{\lambda=\rmT,0}\int\frac{\der^2\bt\der^2\bt'}{(2\pi)^4}  e^{-i\Kt\cdot\rbbpt}\cos\left(n \Phi\right)\hat G^{0}_{\eta_f}(\rbbpt)\Scal(P_\perp,\mu_b)\nonumber\\
    &\times\frac{2\alpha_s(-1)^{p+1}}{n\pi N_c}\left\{\sum_{k=1}^p\frac{1}{k}-\frac{1}{n}-\frac{1}{2}\ln(z_1z_2)\right\}\nonumber\\
    &+\alpha_s\alpha_{\rm em}e_f^2 \int_0^1\frac{\der \zeta_1}{\zeta_1^2}\int_0^1\frac{\der \zeta_2}{\zeta_2^2}D_{h_2/\bar q}(\zeta_2,\mu_F^2)\overline{D}_{h_2/\bar q}(\zeta_2,\rbbpt;\mu_F,P_\perp)\overline{D}_{h_1/q}(\zeta_1,\rbbpt;\mu_F,P_\perp)2\cos(n\phi) \ \nonumber\\
    &\times \ \delta\left(1-z_1-z_2\right)\Hcal_{\rm LO}^{\lambda=\rmT,0}\left(\frac{-2\chi^2}{1+\chi^4}\right)\int\frac{\der^2\bt\der^2\bt'}{(2\pi)^4}  e^{-i\Kt\cdot\rbbpt}\cos\left(n\Phi\right)\hat h^{0}_{\eta_f}(\rbbpt)\Scal(P_\perp,\mu_b)\nonumber\\
    &\times\frac{\alpha_s(-1)^{p}}{(n^2-4)\pi N_c}\left\{(n+2)\sum_{k=1}^{p-1}\frac{1}{k}+(n-2)\sum_{k=1}^{p+1}\frac{1}{k}-\frac{2(n^2+4)}{n^2-4}-n\ln(z_1z_2)\right\} \,.\label{eq:NLO-impact-factor-final2p-T}
\end{align}

\bibliographystyle{utcaps}
\bibliography{dihadron-NLO}

\end{document}